\documentclass[sigconf]{acmart}
\usepackage[utf8]{inputenc}

\usepackage{amsfonts}

\usepackage{amssymb}
\usepackage{subfigure}
\usepackage{graphicx}

\usepackage{amsfonts}
\usepackage{siunitx}

\usepackage{color}

\usepackage{enumitem}

\usepackage{url}
\makeatletter
\def\url@leostyle{%
  \@ifundefined{selectfont}{\def\UrlFont{\sf}}{\def\UrlFont{\small\ttfamily}}}
\makeatother
\makeatletter
\g@addto@macro{\UrlBreaks}{\UrlOrds}
\makeatother
\newcommand{\eat}[1]{}

\usepackage{fancybox}
\usepackage{mathtools}

\usepackage{xcolor}
\definecolor{light-gray}{gray}{0.9}

\newenvironment{packed_enum}{%
  \begin{enumerate}%
  }{\end{enumerate}}

\usepackage{amsthm}
\newtheorem{definition}{Definition}
\newtheorem{theorem}{Theorem}
\newtheorem{lemma}{Lemma}

\newcolumntype{L}[1]{>{\raggedright\let\newline\\\arraybackslash\hspace{0pt}}m{#1}}

\title{How Hard is Takeover in DPoS Blockchains? \\
Understanding the Security of Coin-based Voting Governance}

\begin{document}
\fancyhead{}
%-------------------------------------------------------------------------------

%don't want date printed
% \date{}

% \title{\Large \bf Demystifying Decentralization in DPoS Blockchain on the Web}
% \title{\Large \bf Towards Understanding the Decentralization and Takeover of DPoS: \\
% An Empirical Analysis of Steem and Hive}
% \title{\Large \bf Towards Understanding the Sword of Damocles over DPoS: \\
% An Empirical Analysis of Decentralization and Takeover of Steem}
% \title{\Large \bf The Sword of Damocles over DPoS Blockchains: \\
% Perspectives From Data-Driven Analysis on the Steem Takeover}
% \title{\Large \bf Concentration isn't always bad: An empirical analysis of the security of coin-based voting governance in DPoS blockchains}

% Demystifying the Decentralization and Takeover of DPoS:
% Perspectives From Data-Driven Analysis on Steem

% %for single author (just remove % characters)
% \author{
% {\rm Your N.\ Here}\\
% Your Institution
% \and
% {\rm Second Name}\\
% Second Institution
% % copy the following lines to add more authors
% % \and
% % {\rm Name}\\
% %Name Institution
% } % end author

\author{Chao Li}
\affiliation{%
  \institution{
  % School of Computer and Information Technology \\
  Beijing Key Laboratory of Security and Privacy in Intelligent Transportation \\
  Beijing Jiaotong University}
  \city{Beijing}
  \country{China}
}
\email{li.chao@bjtu.edu.cn}

\author{Balaji Palanisamy}
\authornote{\noindent Corresponding authors: Wei Wang, Balaji Palanisamy and Jiqiang Liu}
\affiliation{%
  \institution{
  Department of Informatics and Networked Systems \\
  School of Computing and Information \\
  University of Pittsburgh}
  \city{Pittsburgh}
  \country{USA}
}
\email{bpalan@pitt.edu}

\author{Runhua Xu}
\affiliation{%
  \institution{School of Computer Science and Engineering \\
  Beihang University}
  \city{Beijing}
  \country{China}
}
\email{runhua@buaa.edu.cn}

\author{Li Duan}
\affiliation{%
  \institution{
  % School of Computer and Information Technology \\
  Beijing Key Laboratory of Security and Privacy in Intelligent Transportation \\
  Beijing Jiaotong University}
  \city{Beijing}
  \country{China}
}
\email{duanli@bjtu.edu.cn}

\author{Jiqiang Liu}
\authornotemark[1]
\affiliation{%
  \institution{
  % School of Computer and Information Technology \\
  Beijing Key Laboratory of Security and Privacy in Intelligent Transportation \\
  Beijing Jiaotong University}
  \city{Beijing}
  \country{China}
}
\email{jqliu@bjtu.edu.cn}

\author{Wei Wang}
\authornotemark[1]
\affiliation{%
  \institution{
  % School of Computer and Information Technology \\
  Beijing Key Laboratory of Security and Privacy in Intelligent Transportation \\
  Beijing Jiaotong University}
  \city{Beijing}
  \country{China}
}
\email{wangwei1@bjtu.edu.cn}

%-------------------------------------------------------------------------------
\begin{abstract}
%-------------------------------------------------------------------------------
Delegated-Proof-of-Stake (DPoS) blockchains, such as EOSIO, Steem and TRON, are governed by a committee of block producers elected via a coin-based voting system.
We recently witnessed the first de facto blockchain takeover that happened between Steem and TRON.
Within one hour of this incident, TRON founder took over the entire Steem committee, forcing the original Steem community to leave the blockchain that they maintained for years. This is \textcolor{black}{a historical} event in the evolution of blockchains and Web 3.0. 
Despite its significant disruptive impact, little is known about how vulnerable DPoS blockchains are in general to takeovers and the ways in which we can improve their resistance to takeovers.

In this paper, we demonstrate that the resistance of a DPoS blockchain to takeovers is governed by both the theoretical design and the actual use of its underlying coin-based voting governance system.
When voters \textcolor{black}{actively cooperate to resist potential takeovers, our theoretical analysis reveals that the current active resistance of DPoS blockchains is far below the theoretical upper bound.}
However in practice, \textcolor{black}{voter preferences could be significantly different.}
This paper presents the first large-scale empirical study of the passive takeover resistance of EOSIO, Steem and TRON.
\textcolor{black}{Our study identifies the diversity in voter preferences and characterizes the impact of this diversity on takeover resistance.}
Through both theoretical and empirical analyses, our study provides novel insights into the security of coin-based voting governance and suggests potential ways to improve the takeover resistance of any blockchain that implements this governance model.
% Our study identifies the phenomenon of voting power decay due to differences in voter preferences and characterizes the impact of diverse voter preferences on takeover resistance.
% , our theoretical analysis of the impact of the design of the underlying voting system on active takeover resistance demonstrates that the current resistance of DPoS blockchains is far below the theoretical upper bound.
% However in practice, most voters in DPoS blockchains do not take active defensive measures and their voting preferences could be significantly different.
\end{abstract}

\begin{CCSXML}
  <ccs2012>
     <concept>
         <concept_id>10002978.10003006.10003013</concept_id>
         <concept_desc>Security and privacy~Distributed systems security</concept_desc>
         <concept_significance>500</concept_significance>
         </concept>
   </ccs2012>
\end{CCSXML}
  
\ccsdesc[500]{Security and privacy~Distributed systems security}

\keywords{Blockchain; Decentralized Governance; Governance Security; Voting Governance; Delegated Proof of Stake; Web 3.0}

\maketitle

\section{Introduction}
\label{sec:intro}
%-------------------------------------------------------------------------------

% \noindent \textbf{motivate dpos.}
Blockchain technologies are fueling the emergence of decentralized applications and Web 3.0,
where authority and power are spread across the network without interference from any single entity.
Traditional Proof-of-Work (PoW) consensus protocols, used by Bitcoin~\cite{nakamoto2008Bitcoin} and Ethereum 1.0~\cite{buterin2014next}, require the decentralized consensus to be made throughout the entire network. As a result, the throughput of transactions in these networks is limited by the network scale (e.g., Bitcoin has a maximum throughput of 7 transactions/sec~\cite{croman2016scaling}), making it practically difficult to satisfy the needs of many applications.
\textcolor{black}{On the other hand, traditional Proof-of-Stake (PoS) consensus protocols, as adopted by blockchains such as Ethereum 2.0~\cite{kim2020ethereum}, require coin holders to make substantial collateral deposits (e.g., 32 ETH in Ethereum) to participate in governance, preventing the involvement of numerous coin holders with insufficient funds.}
To address scalability and participation concerns, 
the Delegated Proof-of-Stake (DPoS) consensus protocol~\cite{larimer2014delegated} has recently gained popularity and has given rise to a series of successful blockchains, such as EOSIO~\cite{he2021eosafe}, Steem~\cite{li2019incentivized} and TRON~\cite{TRON}\footnote[1]{\textcolor{black}{We chose EOSIO, TRON, and Steem for our study due to their representativeness within the DPoS ecosystem~\cite{kwon2019impossibility}, their rich data relevant to takeovers, as well as the widespread attention they have received from researchers~\cite{he2021eosafe,liu2022decentralization,li2019incentivized,guidi2020graph,nguyen2022sochaindb}.}}.
% These three blockchains encompass a wide range of use cases and have significantly influenced the development of DPoS systems.}}.
% ~\cite{TRON}\footnote[1]{EOSIO and TRON were consistently in the Top 10 cryptocurrencies at coinmarketcap.com in 2019, and Steem was in the Top 25 in 2018. The sum of the all-time high records of their market capitalization is \$34 billion.}. 
In DPoS, the consensus is only reached among a small set of block producers (BPs) (e.g., 21 BPs in EOSIO and Steem and 27 BPs in TRON).
%  which makes DPoS one of the most promising technical choices for satisfying the needs of applications requiring high transaction throughput.
\textcolor{black}{Furthermore, any coin holder can participate in BP elections, making DPoS a very promising technical choice for applications that require high transaction throughput and inclusive governance participation.}

\noindent \textbf{Coin-based voting governance.}
DPoS blockchains are governed by a committee of block producers (BPs) who are periodically elected by coin holders\footnote[2]{Blockchains usually issue tradable cryptocurrencies as coins (e.g., EOS for EOSIO, TRX for TRON and STEEM for Steem). }
% Holders of coins issued by the same blockchain can directly trade within the blockchain. Holders of coins issued by different blockchains need to trade through exchanges.} 
via a coin-based voting system.
Coin holders (or voters) are encouraged to stake (i.e., freeze) their coins to gain voting power and cast votes that are weighted by their voting power.
The top candidates ranked by the received voting power then become BPs. In DPoS, BPs are essentially the rule makers of the blockchain.
BPs can update a wide range of rules in the blockchain by executing a proposal, sometimes called a \textit{fork}, ranging from changing system parameters to blacklisting certain accounts, or even reversing confirmed transactions, as long as the supermajority of BPs (15 out of 21 BPs in EOSIO, 17 out of 21 BPs in Steem and 19 out of 27 BPs in TRON) agree on the proposal.
For instance, in TRON, BPs can propose to modify the amount of block generation reward~\cite{TRON_super}, which allows them to determine their own salaries.
% \footnote[3]{\url{https://tronprotocol.github.io/documentation-en/mechanism-algorithm/sr/}}
A more interesting incident occurred in Nov. 2018 when 16 out of 21 BPs of EOSIO approved the first-ever proposal of changing the private key of an EOS account to resolve a dispute on the account's ownership~\cite{ECAF}.
% \footnote[4]{\url{https://eosauthority.com/approvals/view?scope=libertyblock&name=chngkeyha4ta&lnc=z&network=eos}}. 
This marks a significant event in the history of blockchains as an account's private key, used for signing transactions issued by the account is generally considered to be immutable.

\noindent \textbf{Takeover.}
A takeover in DPoS blockchains refers to an attacker controlling the supermajority of BPs and as a result, gaining immense control of the blockchain including the ability to reverse confirmed transactions and change the private keys of accounts.
%Due to its similarity to a hostile takeover, namely the acquisition of one company by another company against the wishes of the former, we call it a takeover attack.
% We have seen tradeoffs between scalability and decentralization in DPoS blockchains, but what about security?
In contrast, the most significant attack in PoW blockchains is the double-spending attack~\cite{nakamoto2008Bitcoin,karame2012double}, which occurs when an attacker with the majority of the mining power reverses confirmed transactions to spend a coin twice.
% Such an attack in DPoS blockchains is quite similar to a hostile takeover, referring to the acquisition of one company by another company against the wishes of the former, so we call it a takeover attack.
The first de facto takeover attack, known as TRON's takeover of Steem, has occurred recently.
In early 2020, TRON founder purchased pre-mined coins\footnote[3]{The amount of coins issued to founders as rewards, which is about 20\% of STEEM total supply in this case.} from Steemit Inc.~\cite{Steemit_Inc}, the company that launched the Steem blockchain.
% \footnote[6]{\url{https://www.steem.center/index.php?title=Steemit,_Inc}}, 
Although Steemit Inc. promised to never use these coins in BP election, TRON did not make such a commitment.
Therefore, the top BPs in Steem (those not belonging to Steemit Inc.) prohibited the use of pre-mined coins in BP election via \textit{fork} 0.22.2~\cite{Fork_0.22.2}.
% \footnote[7]{\url{https://steemit.com/steem/@softfork222/soft-fork-222}}.
However, on Mar. 2, 2020, within one hour, all the BPs in Steem were quickly replaced by accounts controlled by TRON founder, who then immediately revoked \textit{fork} 0.22.2 via \textit{fork} 0.22.5~\cite{Fork_0.22.5}, forcing the original BPs and the Steem community to leave the blockchain they maintained for years.
% \footnote[8]{\url{https://github.com/steemit/steem/pull/3618}} 

TRON's takeover of Steem is not the only attempt of takeovers in DPoS blockchains.
%while extremely severe, may not be unique.
In Dec. 2021, Block.one, the company that launched the EOSIO blockchain, announced its plan of transferring pre-mined coins (about 6\% of EOS total supply) to another company.
At that moment, the top BPs in EOSIO are members of an organization named EOS Network Foundation~\cite{ENF}.
% \footnote[9]{\url{https://eosnetwork.com/}}.
% {\bf Chao, it would be good to clarify what is this community}
Therefore, they deployed a proposal which basically stopped Block.one from controlling pre-mined coins~\cite{ENF_Statement}.
% \footnote[10]{\url{https://eosnetwork.com/blog/enf-statement-on-eos-network-actions-taken-on-december-8th}}.
As a result, takeover did not happen in this instance. The reason is directly attributed to the higher takeover resistance in EOSIO, a key topic of focus in this paper.

\iffalse
In early 2020, TRON founder purchased Steemit Inc., a company founded by Steem founder, and took control of pre-mined coins controlled by Steemit Inc., which started the Steem Takeover Battle summarized in Figure~\ref{front_page}.
There were promises from Steemit Inc. that the pre-mined tokens (about 20\% of STEEM supply), or founder's reward, would be non-voting stake. 
However, at the moment of takeover, details on the usage of pre-mined tokens after the purchase had not been agreed upon.
Therefore, top witnesses decided to enshrine the promises in code by implementing the Fork 0.22.2~\footnote[2]{https://steemit.com/steem/@softfork222/soft-fork-222}, 
which basically restricted the abilities of five relevant accounts (e.g. \textit{@steemit}) in terms of transferring tokens and partaking in witness election.
Later, TRON founder took fast action against 0.22.2.
On March 2nd 2020, within one hour, all the top-20 witnesses were suddenly replaced by newcomers, who then immediately revoked the restriction via the Fork 0.22.5~\footnote[3]{https://github.com/steemit/steem/pull/3618}.
Finally, instead of fighting a losing battle, the original witnesses decided to move to Hive, a blockchain hard forked Steem, excluding the pre-mined tokens.
To sum up, the Steem takeover is the first de facto takeover of decentralized blockchain by centralized capital.
However, little is known about the context of decentralization of Steem before and after the takeover, as well as the crucial behaviors and interactions of major players during the battle.
\fi

\noindent \textbf{This paper.}
Despite its significant disruptive impact, little is known about how vulnerable DPoS blockchains are in general to takeovers and the ways in which we can improve their resistance to takeovers.
In this paper, we demonstrate that the resistance of a DPoS blockchain to takeovers is governed by both the theoretical design and the actual use of its underlying coin-based voting governance system.
We formally describe a three-phase model for coin-based voting governance and formalize the \textcolor{black}{takeover attack and resistance model} based on our analysis of TRON's takeover of Steem.
We formally model the \textit{takeover game} between an attacker and \textcolor{black}{the cooperative resisters} and prove the existence of a Nash equilibrium.
% When voters actively take defensive measures against potential takeovers, our theoretical analysis of the impact of the design of the underlying voting system on takeover resistance demonstrates that the current resistance of DPoS blockchains is far below the theoretical upper bound. 
When voters \textcolor{black}{actively cooperate to resist potential takeovers}, our theoretical analysis of the impact of the design of the underlying voting system on takeover resistance demonstrates that the current active resistance is far below the theoretical upper bound. 
% However in practice, most voters in DPoS blockchains do not take active defensive measures and their voting preferences could be significantly different. Our findings suggest that, in practice, the theoretical upper bound may not be achieved.
However in practice, \textcolor{black}{voter preferences could be significantly different.} 
We present the first large-scale empirical study of the passive takeover resistance of EOSIO, Steem and TRON.
% Our study identifies the phenomenon of voting power decay due to differences in voters' voting preferences and characterizes the impact of voting power decay on takeover resistance.
\textcolor{black}{Our study identifies the diversity in voter preferences and characterizes the impact of this diversity on takeover resistance.}
Through both theoretical and empirical analyses, our study provides novel insights into the security of coin-based voting governance and suggests potential ways to improve the takeover resistance of any blockchain that implements this governance model.

\iffalse
\noindent \textbf{Contributions.}
In a nutshell, this paper makes the following key
contributions:

\begin{itemize}[leftmargin=*]
\item Our work formally models both coin-based voting governance and \textit{takeover game} in DPoS blockchains. Our analysis proves a Nash equilibrium and suggests the system parameters for achieving the theoretical upper bound of takeover resistance for different DPoS blockchains.

\item To the best of our knowledge, our work presents the first large-scale empirical study of the actual takeover resistance of three top DPoS blockchains for a period of two years. Our study identifies a common phenomenon of voting power decay, characterizes the impact of voting power decay on takeover resistance and suggests the settings for maximizing the actual takeover resistance.

\item With both theoretical and empirical analysis, our study provides novel insights into the security of coin-based voting governance and suggests potential approaches to improve the takeover resistance of both existing and future DPoS blockchains.

\end{itemize}
\fi

\noindent \textbf{Organization.}
\textcolor{black}{
We start by introducing the background in Section~\ref{sec:background}.
We model the coin-based voting governance in Section~\ref{sec_model_governance} and formalize the \textcolor{black}{takeover attack and resistance} model in Section~\ref{sec4}.
In Section~\ref{sec:game}, we investigate the \textit{takeover game} and demonstrate the existence of an upper bound for the \textcolor{black}{active} takeover resistance.
In Section~\ref{sec6}, we study the \textcolor{black}{passive} takeover resistance of EOSIO, Steem and TRON.
\textcolor{black}{We suggest potential ways to improve the takeover resistance and discuss the generalization of our analysis in Section~\ref{sec:discussion}.}
We discuss related work in Section~\ref{sec:related_work} and conclude in Section~\ref{sec:conclusion}.}

\section{Background}
\label{sec:background}
%-------------------------------------------------------------------------------
In this section, we introduce various DPoS blockchains, primarily from the perspective of governance. We focus our discussion specifically around EOSIO, Steem and TRON blockchains.
% We then present preliminaries about three voting systems that have been employed by the selected blockchains.
%This paper selects EOSIO, Steem and TRON as research subjects for three main reasons.
These three blockchains were involved in events related to takeovers recently.
%, namely TRON's takeover of Steem and the dispute between EOS Network Foundation and Block.one. 
Also, these blockchains are among the top cryptocurrency projects that have attracted millions of users and collected billions of transactions\footnote[4]{A basic record of user behavior, such as casting a vote or transferring a coin, is named an action/operation/transaction in EOSIO/Steem/TRON, respectively. In the rest of this paper, we refer to them collectively as transactions.} from users~\cite{kwon2019impossibility}.
Their rich data helps validate our results.
Furthermore, the design of coin-based voting governance in these blockchains is consistent and shares several common aspects which help generalize our results.

\subsection{EOSIO}
\label{sec:EOSIO}
EOSIO is a successful DPoS blockchain and its market capitalization was consistently among the top 10 blockchain projects~\cite{kwon2019impossibility}.
Similar to Ethereum~\cite{buterin2014next}, EOSIO supports smart contracts~\cite{wood2014ethereum} with its underlying virtual machine, enabling developers to quickly build decentralized applications (dapps)\footnote[5]{The back-end of dapps runs by BPs of DPoS blockchains in the form of codes named smart contracts.} on the EOSIO platform.
Rapid developments in EOSIO have attracted researchers to study  various aspects of EOSIO including smart contract security~\cite{he2021eosafe}, dapps~\cite{de2021measuring} and decentralization~\cite{liu2022decentralization}.

\noindent \textbf{Governance in EOSIO.}
The design of the governance system here is primarily based on a combination of two voting rules, liquid democracy~\cite{zhang2021power} and multi-winner approval voting~\cite{scheuerman2021modeling}. %which we will formally model in Section~\ref{sec_model_governance}.
\begin{itemize}[leftmargin=*]
  \item \textbf{Liquid democracy:} 
  This voting rule allows a voter to choose between two options:
  (1) cast her votes directly for BP candidates by herself;
  (2) delegate her voting power to a proxy, who may in turn choose between the two options.
  With the first option, a voter's votes would be weighted by her own voting power.
  However, with the second option, multiple voters may form a delegation chain (i.e., everybody except the end voter in the chain chose option two) or a tree (i.e., everybody except the root voter in the tree chose option two), and voting power of the chain (tree) would be aggregated at the end (root) voter, whose votes would be weighted by the aggregated voting power.
  \item \textbf{Multi-winner approval voting:} 
  In this voting rule, a voter is allowed to cast multiple votes (30 votes in EOSIO) with each vote going to a distinct BP candidate.
  Here, each vote of a voter would be weighted by the voter's entire voting power, including her own voting power and any voting power concentrated from delegations.
  By the end of the election cycles, BP candidates are ranked by the voting power they received and a set of top candidates (top 21 in EOSIO) win the election and form a committee.
  From then on, any proposal issued to the committee needs to be approved by at least 15 BPs to get adopted.
\end{itemize}

\subsection{Steem}
\label{sec:Steem}
Steem is another prominent DPoS blockchain that supports numerous social applications.
There have been over 324 Steem-based decentralized applications~\cite{Steem_DAPPs}, many of which are designed to serve social users. 
% \footnote[13]{\url{https://steem.com/developers/}}
\textit{Steemit}~\cite{Steemit} is one of the first and the most prominent application in Steem. It represents a decentralized version of Reddit, where users can create and share content as blog posts to receive replies, reposts, upvotes or downvotes.
% Each upvote or downvote is weighted by the coin owned by the voter.
The platform periodically allocates a number of coins called STEEM to reward authors of top-ranked posts.
% The Steem blockchain stores data generated by Steem-based decentralized applications.
% The blockchain creates a new block every three seconds to record all the operations verified by witnesses.
% In Steem, there have been more than thirty different types of operations. 
%As a very successful social application engine powered by blockchain, recently, 
Steem has received extensive attention from both the blockchain community~\cite{li2019incentivized,li2020comparison} as well as the social network community~\cite{nguyen2022sochaindb,guidi2020graph} in the recent years.

\noindent \textbf{Governance in Steem.}
The governance system in Steem is very similar to that of EOSIO.
% , except for a few changes in the system parameters.
% Similar to EOSIO, 
Steem also employs both liquid democracy and multi-winner approval voting and allows each voter to cast at most 30 votes. Steem is governed by a committee of 21 members.
However, there are two main differences between EOSIO and Steem.
In Steem, only 20 out of 21 BPs are determined by the election, while the last BP in the committee is rotated among candidates outside the top 20.
Also, a proposal in Steem needs to receive 17 approvals to get implemented.

\begin{table}
  \small
  \begin{center}
  \begin{tabular}{|p{0.8cm}|p{1.55cm}|p{1.25cm}|p{1.1cm}|p{1.6cm}|}
  \hline
  {\textbf{Chain}} & {\textbf{Voting Rule}} & {\textbf{MaxVote ($v$)}} & {\textbf{CmteSize ($n$)}} & {\textbf{MinApprov ($t$)}} \\
      \hline
      EOSIO & AV(+LD) & 30 & 21   & 15 \\
      \hline % \hdashline
      Steem & AV(+LD) & 30 & 20(+1) & 17 \\
      \hline % \hdashline 
      TRON  & CV    & 30 & 27   & 19 \\
      \hline
  \end{tabular}
  \end{center}
  \caption{\small Summary of key design choices made by EOSIO, Steem and TRON. Here, LD/AV/CV refer to liquid democracy, approval voting and cumulative voting, respectively.}    
  \vspace{-7mm}    
  \label{table:summary}
\end{table}

\subsection{TRON}
TRON is one of the youngest blockchains employing proof-of-stake principles as its consensus algorithm. Its market capitalization was also among the top-20 blockchain projects~\cite{kwon2019impossibility,coinm}.
Similar to EOSIO, through its support for smart contracts, the ecosystem of TRON has quickly spread across various areas including Non-Fungible Token (NFT), stable coins and decentralized exchanges.
% Decentralized Finance (DeFi), 

\noindent \textbf{Governance in TRON.}
The governance system in TRON is quite different from those in EOSIO and Steem.
TRON replaces liquid democracy and approval voting with cumulative voting~\cite{bhagat1984cumulative}, another well-studied voting system. 
We briefly introduce its concept here and we formally model it in Section~\ref{sec_model_governance}.

\begin{itemize}[leftmargin=*]
  \item \textbf{Multi-winner cumulative voting:} 
  Similar to approval voting, multi-winner cumulative voting allows a voter to cast multiple votes (30 votes in TRON) with each vote going to a distinct BP candidate.
  However, unlike approval voting, here, if a voter decides to cast multiple votes, she must divide her entire voting power into different votes so that the sum of voting power allocated to all votes is no more than her voting power in total.
  Similar to approval voting, BP candidates are then ranked by the voting power they received and multiple top candidates (top 27 in TRON) win the election (i.e., become BPs) and form a committee.
  In TRON, however, a proposal needs to be approved by at least 19 BPs to get adopted.
\end{itemize}

\textcolor{black}{
In Table~\ref{table:summary}, we summarize the key design choices made by EOSIO, Steem and TRON.
Please note that, in the rest of this paper, we denote the max votes per voter parameter by MaxVote $v$, the committee size parameter by CmteSize $n$ and the min approvals per proposal parameter by MinApprov $t$.}

%-------------------------------------------------------------------------------
\section{Coin-based Voting Governance}
\label{sec_model_governance}
%-------------------------------------------------------------------------------
% In this section, we first introduce the background about how coin-based voting governance is working in existing DPoS blockchains represented by EOSIO, Steem and TRON.
% We then extract a general three-phase governance model and summarize differences among EOSIO, Steem and TRON in each phase of the model.
\textcolor{black}{
  In this section, we distill the governance systems introduced in Section~\ref{sec:background} into three distinct phases and provide a formal description of a three-phase model for coin-based voting governance. }

\subsection{Phases of Coin-based Voting Governance}
\label{s3.1}

\textcolor{black}{
  In coin-based voting governance, the process of gradually transforming individual coins into governance decision-making power takes place through three distinct phases. This process enables all coin holders to participate in the process while maintaining a unique balance between scalability and decentralization. }

\begin{itemize}[leftmargin=*]
  \item \textcolor{black}{\textbf{Phase 1: staking.} 
  During the first phase, individual coins are converted into individual voting power. 
  Coin holders lock or \textit{stake} their coins to obtain voting power proportional to the amount of coins staked. 
  This encourages active participation from all coin holders, fostering inclusiveness and decentralization in the decision-making process.
  % During the first phase, individual coins are converted into individual voting power. 
  % Coin holders (or voters) stake (i.e., freeze/lock) their \textit{coins} to gain \textit{voting power}.
  % The amount of voting power acquired is proportional to the number of coins frozen, incentivizing voters to participate in the governance process.
  }
  
  \item \textcolor{black}{\textbf{Phase 2: voting.}
  In the voting phase, individual voting power is aggregated. Coin holders cast their votes, weighted by their voting power, in support of their preferred block producers (BPs). This pooling of voting power enables the community to collectively determine the most suitable BPs for governance.}
  % In the second phase, the individual voting power is pooled together.
  % With their voting power, coin holders cast weighted votes for BP candidates. Candidates who accumulate the highest voting power are elected as BPs and entrusted with the responsibility of managing the blockchain and updating the rules.}
  
  \item \textcolor{black}{\textbf{Phase 3: governing.}
  In the governing phase, the pooled voting power is converted into governance decision-making power. The top candidates ranked by the received voting power become BPs. These BPs form a smaller consensus group that is responsible for making decisions on behalf of the entire network. 
  % This arrangement allows for efficient decision-making and high transaction throughput without compromising the decentralized nature of the blockchain.
  % Elected BPs hold the authority to propose and execute changes in the blockchain, commonly referred to as \textit{forks}. A supermajority agreement among BPs (e.g., 2/3) is typically required for a proposal to be executed, ensuring that the blockchain maintains its decentralized and secure nature.
  }
  
\end{itemize}

% \textcolor{black}{
%   By seamlessly transitioning through these three phases, coin-based voting governance systems effectively balance participation, decentralization, and scalability, making them an attractive choice for modern blockchain networks.
% }

\begin{figure}
  \centering
  {
      \includegraphics[width=1\columnwidth]{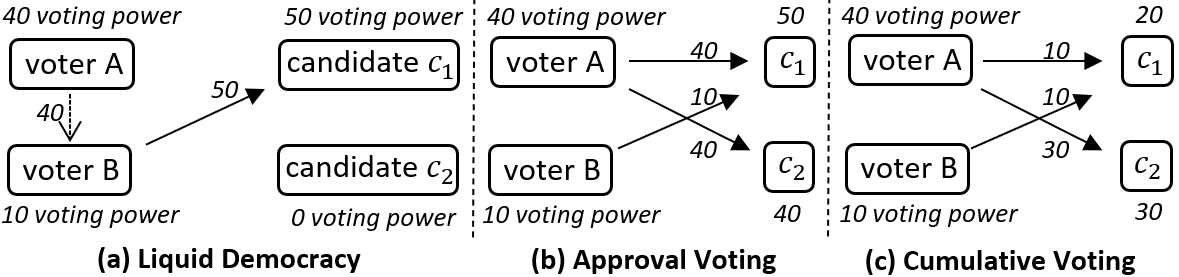}
  }
  \vspace{-7mm}
  \caption {\small Phase 2 with distinct voting systems.}
  \vspace{-4mm}
  \label{phase_2_example} 
\end{figure}

\textcolor{black}{Among them, Phase 2 may utilize distinct voting systems for aggregating individual voting power, resulting in diverse outcomes. }

\noindent \textcolor{black}{\textbf{Example.}
In the example illustrated by Figure~\ref{phase_2_example}, voters A and B obtain 40 and 10 voting power in Phase 1, and vote for candidates $c_1$ and $c_2$ in Phase 2. 
With liquid democracy, voter A delegates her 40 voting power to voter B, increasing B's weight to 50. 
In approval voting, both of A's votes receive a full 40 voting power.
In cumulative voting, A's two votes share her 40 voting power.
}

\subsection{Modeling Coin-based Voting Governance}

\begin{figure}
  \centering
  {
      \includegraphics[width=1\columnwidth]{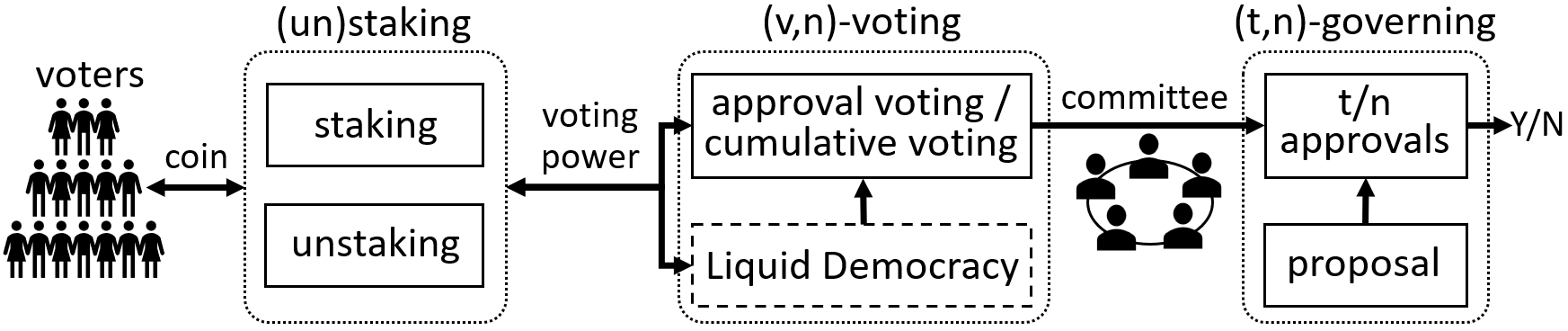}
  }
  \vspace{-6mm}
  % \caption {A three-phase model of coin-based voting governance in DPoS blockchains}
  \caption {\small A model of coin-based voting governance}
  \vspace{-4mm}
  \label{govern_model} 
\end{figure}

We present a high-level overview of the coin-based voting governance model in Figure~\ref{govern_model}.
%  and formally describe the voting rules in the context of DPoS blockchains.
We consider a setting $(M,C)$ for coin-based voting governance based on voting rules~\cite{scheuerman2021modeling}, where $M = \{1,2,...,m\}$ represents the set of voters and $C = \{c_1,...,c_k\}$ represents the set of candidates. Based on this notion, we model the three phases of coin-based voting governance described in Section~\ref{s3.1}.

\subsubsection{\textbf{Phase 1: (un)staking}}

% \noindent \textbf{Phase 1: (un)staking.}
To make any contribution to the governance, a voter $i$ needs to \textit{stake} (i.e., freeze/lock) her coins to earn some voting power via a staking function $p_i = \textit{S}(coin_i, \lambda)$, where $coin_i$ represents the coins of the voter and the parameter $\lambda$ governs how much voting power is earned by each coin. The staking function returns the amount of voting power $p_i$ for the voter. The configuration of the parameter $\lambda$ typically varies across different DPoS blockchains.
In TRON, $\lambda=1$ and it indicates that one coin simply corresponds to one unit of voting power.
In Steem, $\lambda$ is approximately 2000 and therefore, each coin could be converted into 2000 units of voting power.
On the other hand, EOSIO adopts a more sophisticated approach\footnote[6]{We refer the interested readers to~\cite{liu2022decentralization} for more details on EOSIO staking function.} where $\lambda$ is set to be the timestamp of the latest voting transaction performed by the voter. This encourages voters to frequently cast votes to increase the value of $\lambda$ so that they could receive a higher amount of voting power with the same number of staked coins.

In almost all DPoS blockchains, staked coins are not allowed to be withdrawn for a certain period of time, ranging from a few days to several weeks.
After this time, a voter $i$ may choose to unstake her coins using an unstaking function $coin_i = \textit{S}^{-1}(p_i, \lambda)$. Naturally, unstaking the voter's staked coins will result in a decrease of her voting power. Overall, the staking and the unstaking processes results in a dynamically changing voting power profile for the blockchain denoted as $\mathbf{p} = (p_1,...,p_m)$, which captures a snapshot of the voting power of the set of voters $M$ at any given time point.

\subsubsection{\textbf{Phase 2: $(v, n)$-voting}}

% \noindent \textbf{Phase 2: $(v, n)$-voting.}
Given the set of voters and candidates $(M,C)$ and the voting power profile $\mathbf{p}$, the goal of phase 2 is to determine a winning set $W \subseteq C$, also referred to as a committee, which maximizes a community choice score $\tau$.
% \textcolor{black}{namely accumulating the maximum amount of voting power from voters}. 
Phase 2 is referred to as the $(v, n)$-voting phase.
\textcolor{black}{Here, $v$ and $n$ refer to the max votes per voter parameter MaxVote and the committee size parameter CmteSize, respectively, as shown in Table~\ref{table:summary}. 
The score $\tau$ represents the sum of voting power received by the top-$n$ candidates.}
% {\bf Chao, the notion of the community choice score, $\tau$ should be explained here. Also it will be good to briefly remind the meaning of the MaxVote parameter and CmteSize parameter here.}
% the maximum votes per voter and the total number of BPs (i.e., $|W|$), respectively. 
% As can be seen in Table~\ref{table:summary}, different DPoS blockchains may adopt different voting rules.
%In this paper, we focus on the three voting rules adopted by EOSIO, Steem and TRON.

\noindent \textbf{Liquid democracy.}
As discussed in Section~\ref{sec:background}, liquid democracy and approval voting are two primary mechanisms used in the DPoS governance structure.
Specifically, liquid democracy is used in combination with approval voting in both EOSIO and Steem.
% {\bf In fact, as a rule on whether delegating voting power is allowed, liquid democracy could be compatible with various voting rules including approval voting and cumulative voting.} 
On the other hand, a governance system may also choose to not use liquid democracy, as in the case of TRON.
Based on recent work in liquid democracy~\cite{zhang2021power}, we define the delegation profile of the blockchain as $\mathbf{d} = (d_1,...,d_m)$, where $d_i = j$ indicates that voter $i \in M$ is delegating her entire voting power to voter $j \in M$
\footnote[7]{In EOSIO and Steem, a voter is only allowed to delegate her voting power to a single voter and she must delegate her entire voting power.}.
As we discussed in Section~\ref{sec:EOSIO}, voters may form delegation chains (trees), and all their voting power will be aggregated at the end (root) voters, who are commonly referred to as gurus~\cite{zhang2021power}. Note that, if voter $i$ did not delegate her voting power, we have  $d_i = i$ which indicates that voter $i$ is her own guru. 
We define liquid democracy as follows:

\noindent \begin{definition}[\textbf{Liquid Democracy}]
  \textit{Given a triple 
  $\langle M, \mathbf{p}, \mathbf{d} \rangle$ as input, Liquid Democracy determines a couple 
  $\langle M^*, \mathbf{p^*} \rangle$, where $M$ is the set of voters, 
  $\mathbf{p}$ is the voting power profile of voters,
  $\mathbf{d}$ is the delegation profile of voters,
  $M^* \subseteq M$ is the set of gurus,
  $\mathbf{p^*}$ is the (aggregated) voting power profile of gurus.}
\end{definition}

\noindent 
In other words, based on $\langle M, \mathbf{p}, \mathbf{d} \rangle$, liquid democracy would be able to determine a subset $M^*$ of the set of voters $M$ who have either been end (root) voters of delegation chains (trees) or voted by themselves. This subset $M^*$ and its corresponding (aggregated) voting power profile $\mathbf{p^*}$ would then be delivered to the next subphase, namely to the approval voting or cumulative voting subphase.
Please note that, in the rest of this paper, to avoid any confusion between the two terms `voters' and `gurus', we will ignore their differences and use `voters' consistently to refer to the subset $M^*$ because our work is more focused on approval/cumulative voting.

\noindent \textbf{Approval voting.}
We now proceed to modeling approval voting and cumulative voting.
In multi-winner approval voting, \textcolor{black}{multiple winners are determined via approval voting.}
Here, the MaxVote parameter $v$ denotes that a voter $i$ is allowed to cast at most $v$ votes to $v$ distinct BP candidates and each vote is equally weighted by voter $i$'s entire voting power $p^{*}_{i}$.
Therefore, we define a voting profile for all voters in $M^*$  as $V = \{V_i | i \in M^*\}$, 
where voter $i$ selects a subset $V_i$ of the set of candidates $C$ to vote such that $|V_i| \le v$.
We could then define multi-winner approval voting as follows:

\noindent \begin{definition}[\textbf{Multi-Winner Approval Voting}]
  \textit{Given a tuple 
  $\langle M^*, \mathbf{p^*}, V, C, n \rangle$ as input, Multi-Winner Approval Voting determines a committee $W \subseteq C$, such that $|W|=n$ and the community choice score $\tau = \sum_{i \in M^*} |W \cap V_i| p^*_i$ is maximized.}
  % , where $n$ refers to the CmteSize parameter.}
\end{definition}

\noindent 
This definition indicates that after ranking all candidates ($C$) based on the voting power that they have received from all voters ($M^*$), the top $n$ (i.e., CmteSize) candidates form a committee $W$, which is then provided to phase 3. 
\textcolor{black}{The voting power received by the committee $W$ represents the maximized community choice score $\tau$.}
% {\bf Chao, this section makes heavy use of notations. It will be good to describe the notation each time. Fore.g., we could say committee W whenever we say W. It will make it easier for readers to understand without memorizing all the notations. }

\noindent \textbf{Cumulative voting.}
In contrast to approval voting, multi-winner cumulative voting adopts a different approach \textcolor{black}{to determine multiple winners}. Instead of equally weighting each vote by voter $i$'s entire voting power $p^{*}_{i}$, a voter $i$ in cumulative voting has to distribute her voting power $p^{*}_{i}$ across all selected candidates, namely $V_i$.
Thus, we could define a power distribution profile for all voters in $M^*$ as $P = \{P_i | i \in M^*\}$,
where $P_i = \{p^{*}_{i,j} | c_j \in V_i, \sum_j p^{*}_{i,j} \le p^{*}_{i} \}$ so that different candidates $c_j$ selected by voter $i$ may receive different amounts of voting power $p^{*}_{i,j} $ from voter $i$.
We could then define multi-winner cumulative voting as follows:
% The Multi-Winner Cumulative Voting Rule can then be defined as follows:

\noindent \begin{definition}[\textbf{Multi-Winner Cumulative Voting}]
  \textit{Given a tuple 
  $\langle M^*, \mathbf{p^*}, V, P, C, n \rangle$ as input, Multi-Winner Cumulative Voting determines a committee $W \subseteq C$, such that $|W|=n$ and the community choice score 
  $\tau = \sum_{i \in M^*} \sum_{c_j \in W \cap V_i} p^{*}_{i,j}$ is maximized.}
\end{definition}

In summary, multi-winner approval voting allows each unit of voting power to be used for up to $v$ times (i.e., MaxVote), while multi-winner cumulative voting allows each unit of voting power to be used only once.
The output of both voting systems is a committee $W$.
However, the ranking of BP candidates may change whenever new delegating/voting/(un)staking transactions arrive.

\subsubsection{\textbf{Phase 3: $(t, n)$-governing}}

% \noindent \textbf{Phase 3: $(t, n)$-governing.}
Given a committee $W$, in phase~3, every proposal issued to the committee $W$ must receive a minimum of $t$ distinct approvals (i.e., MinApprov in Table \ref{table:summary}) to get adopted.
%The specific configurations of $t$ in EOSIO, Steem and TRON are listed in Table~\ref{table:summary}.

Together, the three phases gradually convert coins owned by holders into voting power to a committee and finally into governance decision-making power.
% The small size of committees in DPoS blockchains significantly increases their transaction throughput.
% Meanwhile, the community choice process helps retain a certain level of decentralization. {\bf Chao, the community choice process is a bit unclear. Further explanation will be helpful} 
However, from the point of view of security, little is known about how vulnerable coin-based voting governance is in general to takeover and the ways in which we can improve their resistance to takeovers.
In the next section, we will start answering questions along these aspects.

\begin{figure}
  \centering
  %\vspace{-3 mm}
  \subfigure[{\small Before TRON's takeover (block 41,297,585)}]
  {
     \label{takeover_snapshot_1}
      \includegraphics[width=0.88\columnwidth]{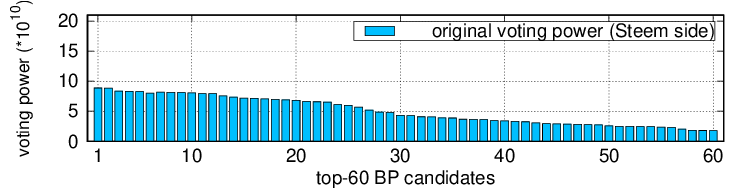}
  }
  \subfigure[{\small After TRON's takeover (block 41,297,734)}]
  {
    \label{takeover_snapshot_2}
      \includegraphics[width=0.88\columnwidth]{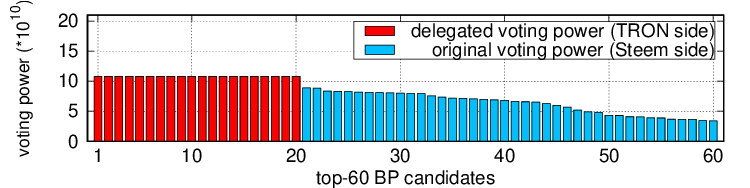}
  }
  \subfigure[{\small After \textit{fork} 0.22.5 (block 41,297,909)}]
  {
    \label{takeover_snapshot_3}
      \includegraphics[width=0.88\columnwidth]{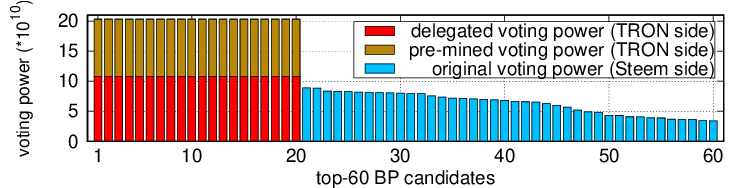}
  }
  \vspace{-5mm}
  \caption{\small The shift in rankings of the top BP candidates in Steem before and after TRON's takeover.}
  \vspace{-3mm}
  \label{takeover_snapshots}
\end{figure}

\begin{figure}
  \centering
  {
      \includegraphics[width=0.88\columnwidth]{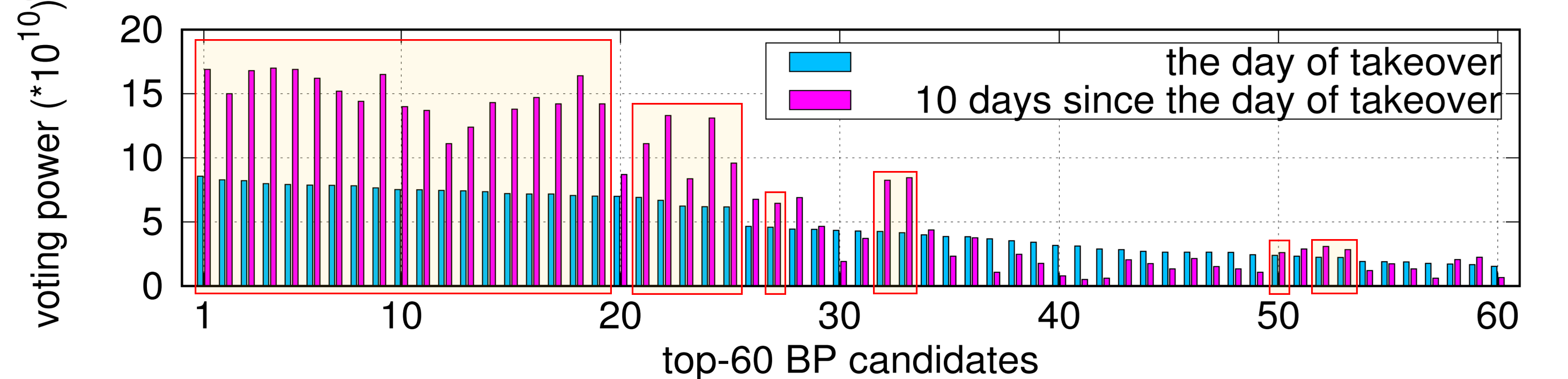}
  }
  \vspace{-4mm}
  \caption {\small \textcolor{black}{The variation in voting power of top BP candidates before takeover versus 10 days after takeover. The boxed BPs represent those suggested by the call-to-action.}}
  \vspace{-3mm}
  \label{takeover_10days} 
\end{figure}

%-------------------------------------------------------------------------------
\section{Takeover Attack/Resistance: Lessons from TRON's takeover of Steem}
\label{sec4}
%-------------------------------------------------------------------------------
\textcolor{black}{
In this section, by reviewing the intricacies of TRON's takeover of Steem and Steem's resistance against TRON's takeover, we introduce and formalize the takeover attack and resistance model. 
We also present the two key research questions that drive our study in the next two sections.}
% Then, after presenting the threat model and attack procedure of takeover, 

\subsection{TRON's Takeover of Steem}
\label{s4.1}
We carefully review the blockchain data of Steem that was generated on the takeover day (March 2, 2020).
% ~\footnote[14]{Clarification: In this paper, we only present the objective information that we obtained from blockchain data. For more details of the event, one may refer to a comprehensive summary: \url{https://decrypt.co/38050/steem-steemit-tron-justin-sun-cryptocurrency-war}. However, we did not verify any addtional information provided by this article and any subjective view of the author does not represent our opinion.}.
Based on our investigation, it is interesting to find out that the takeover was implemented within 44 minutes, from block 41,297,060 to 41,297,909 (3 seconds per block), in three phases that closely follow the model presented in Section~\ref{sec_model_governance} well. It is important to note that we only present the objective information that we obtained from the blockchain data unless explicitly stated.

% {\bf Chao, can we specifically include phase 1, phase 2 and phase 3 in the following paragraphs - it will provide better connection with our model}

\begin{itemize}[leftmargin=*]
\item \textbf{Phase 1: staking.} 
During the first 27 minutes of the 44 minutes, we find that three accounts converted \$7,469,573 worth of coins to voting power.
Meanwhile, based on the Liquid Democracy rule, these three accounts, along with six other accounts, delegated their voting power to the same proxy, which means that any vote cast by this proxy would be weighted by the huge amount of voting power delegated to it. 
% The huge amount of voting power was initially distributed across a few accounts but was later delegated to a single proxy by these accounts based on the Liquid Democracy Rule.
As illustrated in Figure~\ref{takeover_snapshot_1}, by the end of this phase, the proxy had not yet cast any vote and therefore, all the top-60 BP candidates were still controlled by the original voting power owned by the Steem community.

\item \textbf{Phase 2: $(30,21)$-voting.}
During the next 8 minutes after the first 27 minutes, the proxy cast votes to 20 distinct BP candidates one by one. These BP candidates had received nearly no voting power previously. As illustrated in Figure~\ref{takeover_snapshot_2}, by the end of this phase, all the top-20 seats were occupied by the BP candidates that were supported by the proxy.
We can also observe that all the original top BP candidates in Figure~\ref{takeover_snapshot_1}, 
whose voting power did not significantly change during the 8 minutes, 
dropped exactly 20 places in the ranking.
It is worth noting that, as we have introduced in Section~\ref{sec:Steem}, one may at most take over 20 out of all the 21 seats in the committee of Steem because the last seat rotates among candidates outside the top 20.

\item \textbf{Phase 3: $(17,21)$-governing.}
% {\bf Chao, it will be good to remind what fork is - the readers might not remember it}
\textcolor{black}{
During the last 9 minutes, the top-20 BPs had the ability to pass any proposal they wanted.
Recall that the use of pre-mined coins (i.e., the coins that TRON founder purchased from Steemit Inc.) in BP election had been prohibited at an earlier time by a proposal passed by the original committee, namely \textit{fork} 0.22.2.
The new committee then revoked the prohibition of the use of pre-mined coins in BP election by implementing a new proposal, namely  \textit{fork} 0.22.5.}
% TRON founder purchased pre-mined coins~\footnote[5]{The amount of coins issued to founders as rewards, which is about 20\% of STEEM total supply in this case.} from Steemit Inc.
% revoked \textit{fork} 0.22.2 via a new \textit{fork} 0.22.5, which actually revoked the prohibition of the use of pre-mined coins in BP election.
%, as we have introduced in Section~\ref{sec:intro}.
% {\bf Chao, this point about pre-mined coins may be clarified further.}
The pre-mined coins were then immediately used to support the top-20 BPs.
By the end of this phase (Figure~\ref{takeover_snapshot_3}), fuelled by the power of pre-mined coins, all the top-20 BPs gained significant advantages, rendering them nearly undefeated.

\end{itemize}

% \subsection{Steem's Organized Resistance}
\subsection{Steem's Resistance Against TRON's takeover}
\label{s4.2}

\textcolor{black}{
We identified two resistance patterns against TRON's takeover.
}

\noindent 
\textcolor{black}{
% \textbf{The collaborative resistance pattern.}
\textbf{Passive resistance.}
In the takeover process outlined in Section~\ref{s4.1}, the original voting power acquired by BP candidates from Steem community members forms a passive resistance against the takeover, compelling the attacker to amass substantial voting power. For example, during Phase 1, TRON founder accumulated \$7,469,573 worth of coins as voting power.
}

\noindent 
\textcolor{black}{
\textbf{Active resistance.}
Subsequently, we investigated the blockchain data of Steem within ten days after the takeover and discovered an active resistance against the takeover.  
The active resistance consists of two crucial stages: a leader initiates a call-to-action~\cite{call-to-action}, followed by the collaborative response of community members.
More concretely, amidst a hostile takeover, a well-respected community member leads the resistance by issuing a call-to-action, which functions as a rallying cry that inspires the community to protect their shared interests against the takeover attempt.
In response, some community members pool their resources and form a cohesive voting front to counter the takeover. 
Together, they create a formidable voting power dedicated to supporting a list of candidates suggested by the call-to-action.
We formalize the active resistance in Section~\ref{s4.3} and provide more details in Appendix~\ref{appendix_A}.
% Subsequently, we investigated the Steem blockchain data within ten days after the takeover and discovered an organized  resistance pattern employed by the Steem community against the takeover.  
% We offer a brief summary of our findings, with detailed information to be presented in Appendix \ref{app:steem_takeover}.
% We offer a concise overview of the collaborative resistance pattern, with in-depth details provided in Appendix \ref{appendix_A}.
}

\textcolor{black}{
Figure~\ref{takeover_10days} illustrates the practical impact of the active resistance in the case of TRON's takeover. On the day of the takeover, a renowned Steem community member posted a call-to-action~\cite{call-to-action} on the Steemit platform, which garnered the highest number of comments within ten days. Ten days later, all the BPs suggested by the call-to-action witnessed positive growth in voting power and occupied the top 25 rankings, emerging as the core BPs countering the takeover. In contrast, the majority of BPs not endorsed in the call-to-action experienced a decrease in their voting power.}

\subsection{Modeling Takeover Attack and Resistance}
\label{s4.3}
% \subsection{Takeover Attack Model}
% \subsection{Threat Model and Attack Procedure}

\begin{figure}
  \centering
  {
      \includegraphics[width=1\columnwidth]{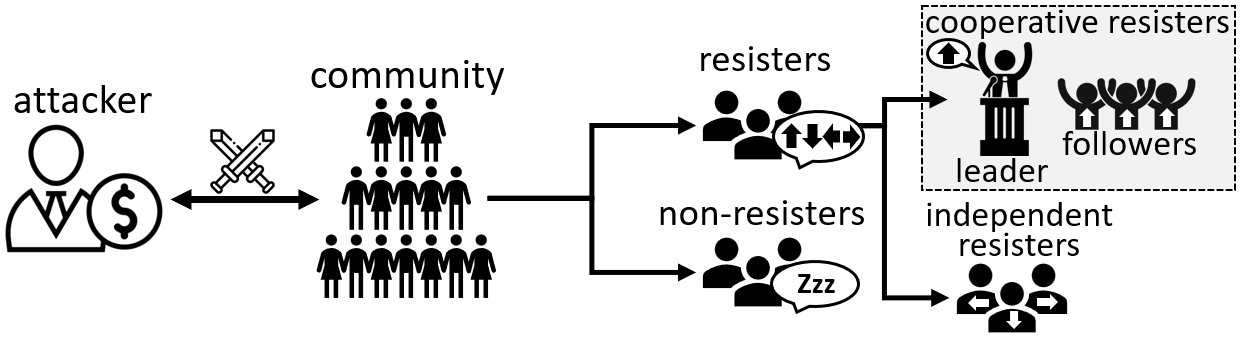}
  }
  \vspace{-8mm}
  \caption {\small The takeover attack and resistance model.}
  \vspace{-4mm}
  \label{war_model} 
\end{figure}

\textcolor{black}{
As illustrated by TRON's takeover of Steem, in a takeover event, an attacker attempts to take over a blockchain, while some community members of the target blockchain strive to resist this takeover. }
% by wresting control of the takeover committee
% It is important to note that the community members are not a monolithic group.
% They can be further divided into resisters and non-resisters based on their level of engagement in the resistance. 
% Moreover, the resisters can be further categorized into cooperative resisters and independent resisters, depending on their votes during the resistance process. 
The takeover attack and resistance model is depicted in Figure~\ref{war_model}.

\noindent 
% \textbf{Attacker.}
\textbf{Attack.}
% We formalize the takeover attack model based on our analysis of TRON's takeover of Steem. 
In a takeover event, the goal of an attacker $\mathcal{A}$ is to be able to launch \textit{forks} (i.e., passing proposals to change the blockchain rule), which is equivalent to occupying at least $t$ seats of the committee.
To do that, an attacker $\mathcal{A}$ needs to explicitly or implicitly control three types of resources, namely a subset $C_a$ of the set of candidates $C$ where $|C_a| \ge t$ (i.e., MinApprov), a subset $M_a$ of the set of voters $M$, as well as an amount of voting power $p_{a}$.
Then, $\mathcal{A}$ needs to follow a strategy $s_a$, which is simply defined as the way of distributing $\mathcal{A}$'s voting power $p_{a}$ across $\mathcal{A}$'s candidates $C_a$ via votes cast by $\mathcal{A}$'s voters $M_a $.
\textcolor{black}{Here, it's worth noting that some voting systems (e.g., approval voting) allow each unit of voting power to be used multiple times. Therefore, we introduce a power amplification coefficient $\zeta_a$ to capture the amplification effect of the voting system settings to $\mathcal{A}$'s voting power $p_a$. The value of $\zeta_a$ is only related to parameters $(v,t,n)$. We present more details of the amplification effect in Section~\ref{sec:game}.}

We formalize takeover attacks as follows:

\noindent \begin{definition}[\textbf{Takeover Attack}]
  \label{def:takeover_attack}
  \textit{An attacker $\mathcal{A}$, 
  who controls $\langle M_a, C_a, p_{a} \rangle$,
  implements a strategy 
  $s_a = \{p_{a,i} | c_i \in C_a, \textcolor{black}{p_{a,i}\leq p_a}, \\ \sum_i p_{a,i} \le \zeta_a p_{a} \}$ of distributing $\zeta_a p_a$ across $C_a$,
  such that the committee $W$ output from the $(v,n)$-voting phase satisfies $|W \cap C_a| \ge t$,
  where $\zeta_a$ is the power amplification coefficient of $\mathcal{A}$.}
\end{definition}

% $p_{a,i}\leq p_a for\ each\ i$ 

\iffalse
\noindent 
In this definition, $\zeta_a$ is used to capture the amplification effect (i.e., reuse each unit of voting power multiple times in approval voting) of the voting system settings to $p_a$, which is only related to parameters $(v,t,n)$. We present more details of the amplification effect in Section~\ref{sec:game}.
% Also, we provide a simplified definition of strategy $s_a$ here which we elaborate further in Section~\ref{sec:game}.

In summary, multi-winner approval voting allows each unit of voting power to be used for up to $v$ times (i.e., MaxVote), while multi-winner cumulative voting allows each unit of voting power to be used only once.

Intuitively, depending on whether
a voting system allows voters to weight multiple votes using the
same coins, an approval voting system tends to amplify voters’
power by �� (i.e., MaxVote) while a cumulative voting system does
not.
\fi

\noindent 
\textcolor{black}{
% \textbf{Resisters.}
\textbf{Resistance.}
% It is important to note that in a takeover event, the goals and behaviors of the target blockchain community members may be diverse and complex, rather than being monolithic.
It is important to note that in a takeover event, the behaviors of the target blockchain community members may not be monolithic.
Intuitively, some community members may engage in active resistance, while others may remain indifferent and abstain from taking any action. These two types of community members are referred to as \textit{resisters} and \textit{non-resisters}, respectively. 
% From the Steem community resistance case introduced in Section 4.2, we can see that 
Moreover, the behaviors of resisters may exhibit variations. Some resisters might follow a call-to-action and concentrate their voting power on a few suggested candidates, while others might disregard any suggested candidates. We denote these two type of resisters as \textit{cooperative resisters} and \textit{independent resisters}, respectively.
% We formalize the different types of resisters as follows:
}

\textcolor{black}{
More formally, we categorize community members who modify their selected candidate set by executing delegating/voting transactions within a short period (e.g., 1 day) after the takeover as \textit{resisters}, and those who retain their selected candidate set as \textit{non-resisters}.
Expanding upon this classification, we identify the author of the most popular call-to-action post (e.g.,~\cite{call-to-action}) as the leader and denote the leader's chosen candidate set as $C_{l}$.
We then classify \textit{resisters} with a chosen candidate set $C_{r}$ satisfying $|C_r \cap C_l| \ge 1$ as \textit{cooperative resisters (co-resisters)}, who follow the active resistance pattern introduced in Section~\ref{s4.2}, and those with a $C_{r}$ satisfying $|C_r \cap C_l| = 0$ as \textit{independent resisters (ind-resisters)}.
}

\begin{figure}
  \centering
  {
      \includegraphics[width=0.9\columnwidth]{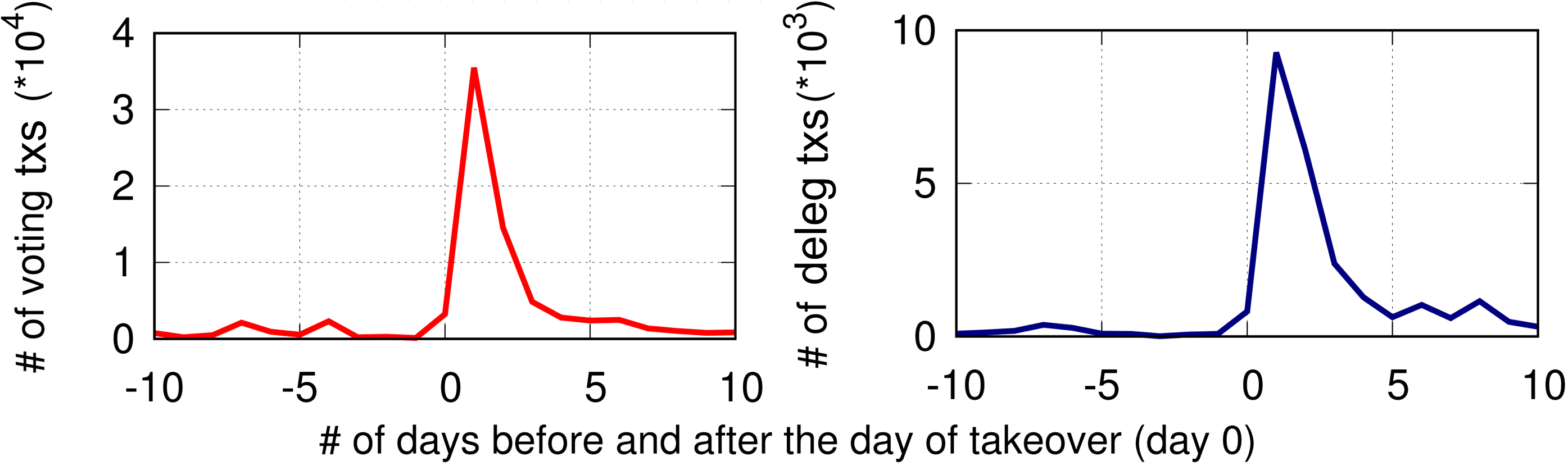}
  }
  \vspace{-3mm}
  \caption {\small \textcolor{black}{Variations in the number of voting transactions (voting txs) and delegating transactions (deleg txs).}}
  \vspace{-4mm}
  \label{ops_variation} 
\end{figure}

\begin{figure}
  \centering
  {
      \includegraphics[width=0.9\columnwidth]{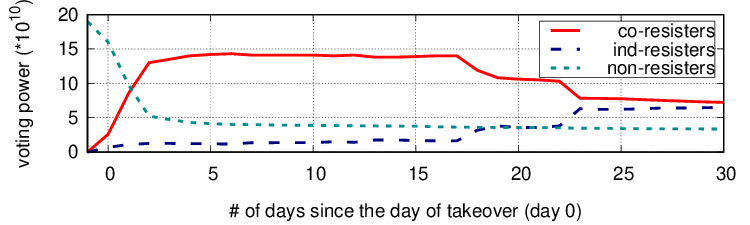}
  }
  \vspace{-3mm}
  \caption {\small \textcolor{black}{Variations in the total voting power of different types of community members.}}
  \vspace{-3mm}
  \label{vp_variation} 
\end{figure}

\textcolor{black}{
In practice, we observe that co-resisters in the Steem community, who adopt the active resistance pattern, serve as the primary force in countering TRON's takeover. 
As depicted in Figure~\ref{ops_variation}, within one day after the takeover, voters generated nearly 40,000 voting/delegating transactions, which is a hundred times the daily average before the takeover. This indicates that a substantial number of community members began to take action. As demonstrated in Figure~\ref{vp_variation}, after the takeover, the total voting power of co-resisters grew rapidly, exceeding ten times the total voting power of ind-resisters on the second day. It then remained stable and surpassed the total voting power of non-resisters by three times on the fifth day, only to decrease when the Steem community began migrating to the new blockchain, Hive~\cite{nguyen2022sochaindb}. This demonstrates that co-resisters played a pivotal role in resisting the takeover.}

Based on the aforementioned analysis, we formalize the active resistance led by co-resisters as follows:

\noindent \begin{definition}[\textbf{Active Resistance}]
  \label{def:active_resistance}
  \textcolor{black}{
  \textit{A group of co-resisters $\mathcal{R}$, 
  who controls an amount of voting power $p_{r}$,
  implements a strategy 
  $s_r = \{p_{r,i} | c_i \in C_l, \textcolor{black}{p_{r,i}\leq p_r}, \sum_i p_{r,i} \le \zeta_r p_{r} \}$ of distributing $\zeta_r p_r$ across the leader's chosen candidate set $C_l$,
  such that the committee $W$ output from the $(v,n)$-voting phase satisfies $|W \cap C_a| < t$,
  where $\zeta_r$ is the power amplification coefficient of $\mathcal{R}$.
  \textcolor{black}{The \textit{active takeover resistance}, denoted as $R_A$, is quantified as the minimum amount of voting power $p_a$ an attacker $\mathcal{A}$ needs to defeat the co-resisters and successfully take over the target blockchain.}}}
\end{definition}

\textcolor{black}{
In this definition, $\zeta_r$ is used to capture the amplification effect (i.e., reuse each unit of voting power multiple times in approval voting) of the voting system settings to $\mathcal{R}$'s voting power $p_r$, which is solely related to parameters $(v,t,n)$. We elaborate more on both $\zeta_r$ and $\zeta_a$ in Section~\ref{sec:game}.
Also, we provide the quantification for the active takeover resistance $R_A$ here and will discuss its theoretical upper bound in Section~\ref{sec:game} in more detail.}

% Based on the above discussion, we can now \textcolor{black}{quantify the \textit{active takeover resistance} as the value of $p_a$ in the equilibrium, denoted as $R_A$}.
% Next, in Lemma~\ref{lemma:3}, we quantify $R_A$ by combining both the two strategies $\widehat{s_a}$ and $\widehat{s_r}$ in the equilibrium.
% Furthermore, we demonstrate the upper bound of $R_A$.

\textcolor{black}{Similarly, when co-resisters are either absent or their power is insignificant, we define the passive resistance as follows:}

\noindent \begin{definition}[\textbf{Passive Resistance}]
  \label{def:passive_resistance}
  \textcolor{black}{
  \textit{The target blockchain community members distribute their voting power across the candidate set $C$.
  The \textit{passive takeover resistance}, denoted as $R_P$, is quantified as the minimum amount of voting power $p_a$ an attacker $\mathcal{A}$ needs to defeat the target blockchain community members and successfully take over the target blockchain.}}
\end{definition}

\subsection{Discussion}

% \noindent \textbf{Active takeover resistance:} 

% \noindent \textbf{Passive takeover resistance:} 

% From TRON's takeover of Steem, we observe two factors that may influence the vulnerability/resistance of DPoS blockchains to a takeover attack:
\textcolor{black}{
From the takeover event between TRON and Steem, we observe two key factors that may influence the active and/or passive resistance.}

\noindent \textbf{Design of the voting system:} 
The first potential factor involves choosing between approval and cumulative voting, as well as selecting parameters $(v,t,n)$.
For instance, intuitively, TRON's takeover would have been more difficult if the Steem blockchain had adopted a smaller $v$ (i.e., MaxVote). 
Currently, in EOSIO, Steem and TRON, the MaxVote parameter $v$ is larger than the MinApprov parameter $t$, enabling an attacker $\mathcal{A}$ to reuse $\mathcal{A}$'s voting power $p_{a}$ to contest each top-20 committee seat, as illustrated by TRON's takeover of Steem.
However, it may be non-trivial to draw conclusions because a smaller $v$ would also constrain the power of both sides.
% {\bf Chao, suddenly $v$ is unclear above though it is explained in the next line - notations may be used along with the actual name of the parameter.}

\noindent \textbf{Actual voter preferences:} 
% The second potential factor refers to the characteristics of a voter's voting preferences, including the number of votes that a voter may cast and the priorities that a voter may assign to her chosen candidates.
The second potential factor pertains to the characteristics of voter preferences, including the number of votes cast and the priorities assigned to selected candidates.
As shown in Figure~\ref{takeover_snapshot_1}, during TRON's takeover, a significant portion of the original voting power was allocated to low-ranking BP candidates due to the diversity of voter preferences.
The phenomenon may be desirable from the perspective of community choice.
However, such a dispersion of defensive voting power may make a DPoS blockchain more vulnerable to takeovers because an attacker's voting power is presumed to be always highly concentrated.

To further analyze active and passive takeover resistance, we pose the following two key research questions. We address each in the subsequent sections.

% \noindent \textbf{RQ 1: }
% When voters actively take defensive measures against potential takeovers, what would be the impact of the setting of parameters $(v,t,n)$ on the resistance of DPoS blockchains to takeovers?

% \noindent \textbf{RQ 2: }
% What do voters' actual voting preferences look like and how the actual voting preferences may affect the resistance of DPoS blockchains to takeovers?

% \begin{mdframed}[innerleftmargin=1.6pt]
%   \begin{packed_enum}[leftmargin=*]
%     \noindent \textbf{RQ 1 [Active Resistance]: }
%     When resistance is led by co-resisters (e.g., Section~\ref{s4.2}), how can the voting system be designed to maximize the effectiveness of active resistance?

%     \noindent \textbf{RQ 2 [Passive Resistance]: }
%     When resistance is passive (e.g., Section~\ref{s4.1}) or the power of co-resisters is much lower than that of non-resisters, 
%     how can we understand actual voter preferences and based on them, how can we design a voting system to enhance the effectiveness of passive resistance?
%   \end{packed_enum}
% \end{mdframed}

\noindent\fbox{%
\parbox{\dimexpr\linewidth-2\fboxsep-2\fboxrule}{
  \begin{enumerate}[leftmargin=*,label={}]
    \item \textcolor{black}{
      \textbf{RQ 1 [Active Resistance]: }
      When resistance is led by co-resisters (e.g., Section~\ref{s4.2}), how can the voting system be designed to maximize the effectiveness of active resistance?}

    \item \textcolor{black}{
      \textbf{RQ 2 [Passive Resistance]: }
      When resistance is passive (e.g., Section~\ref{s4.1}) or the power of co-resisters is much lower than that of non-resisters, 
      how can we understand actual voter preferences and based on them, how can we design a voting system to enhance the effectiveness of passive resistance?}
  \end{enumerate}
}}

% Next, we answer the two research questions in Section~5 and Section~6, respectively.

% For instance, if we break the first condition by changing $(v,t,n)$ to $(4,5,8)$, for voters, the best defensive measure is two evenly divide their voting power into two parts.
% However, now an attacker only needs to 

% \begin{figure}
%   \centering
%   {
%       \includegraphics[width=1.0\columnwidth]{./figures/dpos_pow.png}
%   }
%   % \vspace{-3mm}
%   \caption {PoW 51\% vs. DPoS takeover}
%   % \vspace{-2mm}
%   \label{dpos_pow} 
% \end{figure}

%-------------------------------------------------------------------------------
\section{Active Resistance: Takeover Game}
\label{sec:game}
%-------------------------------------------------------------------------------
\textcolor{black}{In this section, we address the first research question by modeling a \textit{takeover game} between two players, namely an attacker and the cooperative resisters (co-resisters)}.
%  namely a set of voters who actively take defense measures against potential takeovers.
We show the strategies of the two players in a Nash equilibrium and demonstrate the existence of an upper bound for the active takeover resistance of DPoS blockchains for both approval voting and cumulative voting.

\begin{figure}
  \centering
  {
      \includegraphics[width=0.9\columnwidth]{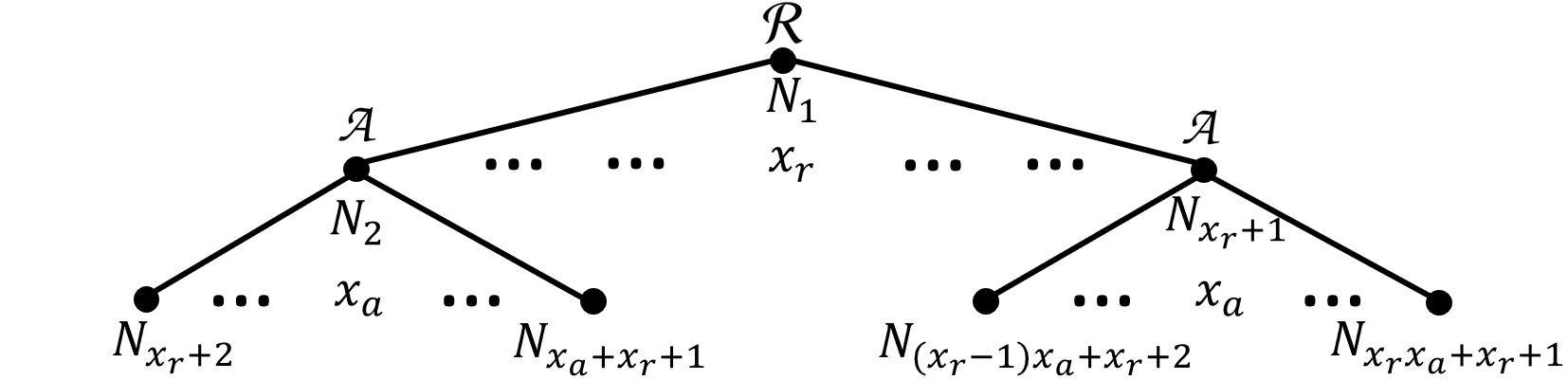}
  }
  \vspace{-3.5mm}
  \caption {\small The game tree in active resistance.}
  \vspace{-4mm}
  \label{game_tree} 
\end{figure}

\noindent \textbf{The game model.}
A takeover attack in DPoS blockchains can be modeled as a perfect-information extensive-form game~\cite{leyton2008essentials},
\textcolor{black}{which reflects the real-world event of TRON's takeover of Steem.} 
The game involves two players $(\mathcal{A},\mathcal{R})$.
The first player is an attacker $\mathcal{A}$ who controls an \textit{alterable} amount of voting power $p_a$ and a set of candidates $C_a$, where $|C_a|=n$ (i.e., CmteSize).
The second player is the co-resisters $\mathcal{R}$ who controls a \textit{fixed} amount of voting power $p_r$ and a set of leader's chosen candidates $C_l$, where $|C_l|=n$ and $|C_l \cap C_a|=0$.
\textcolor{black}{The attacker and co-resisters play sequentially in the game, observing all prior steps from the blockchain. This aligns with the perfect-information extensive-form game model. 
The game concludes in one round because the attacker, controlling an alterable amount of voting power $p_a$, can secure over $t$ seats in the committee (Definition~\ref{def:takeover_attack}). }
The game consists of two stages.
In the first stage, $\mathcal{R}$ needs to determine the strategy $s_r$ of distributing $\mathcal{R}$'s (amplified) voting power $\zeta_r p_r$ across $\mathcal{R}$'s candidates $C_l$.
In the second stage, after learning the distribution of $\mathcal{R}$'s voting power $\zeta_r p_r$ from the blockchain data (i.e, perfect information), $\mathcal{A}$ needs to determine both the required amount of $p_a$ and the strategy $s_a$ of distributing $\mathcal{A}$'s (amplified) voting power $\zeta_a p_a$ across $\mathcal{A}$'s candidates $C_a$.
% Recall that, $\zeta_r$ ($\zeta_a$) captures the amplification effect to $p_r$ ($p_a$) and only depends on $(v,t,n)$, which are assumed to be fixed in the game.
Recall that both $\zeta_r$ and $\zeta_a$ capture the amplification effect to $p_r$ and $p_a$ respectively, and solely depend on $(v,t,n)$, which are constants in this game.
We skip their explanation here as they have no influence on the game.
Next, we express the smallest unit of voting power as $\delta$.
We can then express the number of strategies of distributing $\frac{\zeta_r p_r}{\delta}$ ($\frac{\zeta_a p_a}{\delta}$) pieces of $\delta$ across candidates in $C_l$ ($C_a$) as a finite positive integer $x_r$ ($x_a$). 
Consequently, in this game, $\mathcal{R}$ has a number of $x_r$ (pure) strategies and $\mathcal{A}$ has a number of ${x_a}^{x_r}$ (pure) strategies, as shown in a game tree in Figure~\ref{game_tree}.
% We denote the payoffs of $\mathcal{R}$ and $\mathcal{A}$ as $u_r$ and $u_a$, respectively.
In this game, the goal of $\mathcal{A}$ is to take over the blockchain successfully while minimizing the required voting power $p_a$ and the goal of $\mathcal{R}$ is thus to maximize $p_a$.
\textcolor{black}{
In other words, a higher value of $p_a$ indicates that an attacker $\mathcal{A}$ needs to invest more voting power to defeat the co-resisters $\mathcal{R}$, which suggests higher attack cost and thus higher resistance to takeovers.}
% Intuitively, $\mathcal{R}$ would want to increase $p_a$ while $\mathcal{A}$ would like to make $p_a$ as small as possible.
Therefore, by denoting the payoffs of $\mathcal{R}$ and $\mathcal{A}$ as $u_r$ and $u_a$ respectively, we simply define $u_r=\zeta_a p_a, u_a=-\zeta_a p_a$ for a successful takeover while $u_r=\infty, u_a=-\infty$ denotes a failed one.

% determines a strategy $\Phi_d$ of distributing $p_r$ across candidates in $C_l$.
% After that, $\mathcal{A}$ determines a strategy $\Phi_a$ of distributing $p_a$ across candidates in $C_a$.

\noindent \textbf{The equilibrium.}
Next, we compute the subgame-perfect Nash equilibrium of the game via Backward Induction~\cite{leyton2008essentials},
\textcolor{black}{which proves that the game can rapidly reach equilibrium within a single round.}
The game has $x_r+1$ subgames rooted at the non-leaf nodes, namely $N_1$ to $N_{x_r+1}$ as shown in Figure~\ref{game_tree}.

\textcolor{black}{
Before further analysis, we introduce $p^r_{n-t+1}$, which represents the voting power of the $(n-t+1)^{th}$ candidate in the sorted vector $\mathbf{c^r} = (c^r_1,...,c^r_n)$ of $\mathcal{R}$'s candidates, sorted by their voting power assigned by $\mathcal{R}$ from high to low. Then, based on the introduced $p^r_{n-t+1}$}, we define two key strategies:

\begin{itemize}[leftmargin=*]
  \item \textbf{$\widehat{s_a}$}:
  A strategy of $\mathcal{A}$ in which $\mathcal{A}$ evenly distributes $\zeta_a p_a$ across $t$ candidates in $C_a$.
  \textcolor{black}{For any $s_r$, in the subgame induced by $s_r$, $\zeta_a p_a$ is set to $t p^r_{n-t+1}$ as per the proof of Lemma~\ref{lemma:1} for that subgame.}

  \item \textbf{$\widehat{s_r}$}:
  A strategy of $\mathcal{R}$ in which $\mathcal{R}$ evenly distributes $\zeta_r p_r$ across $(n-t+1)$ candidates in $C_l$.
\end{itemize}

\textcolor{black}{
Intuitively, $\widehat{s_a}$ suggests that an attacker $\mathcal{A}$ should always choose to invest all the voting power $p_a$ just to a minimum number of candidates required by $t$ (i.e., MinApprov) and make these $t$ candidates equally strong so that none of them may be easily defeated by the co-resisters $\mathcal{R}$ as a breakthrough.
Similarly, $\widehat{s_r}$ suggests that the co-resisters $\mathcal{R}$ should always choose to invest all the voting power $p_r$ just to a number of $(n-t+1)$ candidates and make these candidates equally strong so that an attacker $\mathcal{A}$, after easily controlling $t-1$ seats where no defensive power exists, feels difficult to defeat any of the $\mathcal{R}$'s candidates to control the last seat required by $t$. 
}

Next, we prove Theorem~\ref{theorem:1} through two lemmas.

\begin{theorem}
  \label{theorem:1}
  $(\widehat{s_a},\widehat{s_r})$ is the subgame-perfect Nash equilibrium.
\end{theorem}

\begin{lemma}
\label{lemma:1}
  $\widehat{s_a}$ \textcolor{black}{is the unique best response} in any subgame rooted among nodes $N_2$ to $N_{x_r+1}$.
\end{lemma}

\begin{proof}
  \textcolor{black}{In any of these subgames, $\mathcal{A}$ is the sole player in the one-shot game and is given the sorted vector $\mathbf{c^r}$ of $\mathcal{R}$'s candidates.}
  \textcolor{black}{The best response for $\mathcal{A}$ to maximize its payoff $u_a$ is to set $\zeta_a p_a=t p^r_{n-t+1}$ and assign}
  % To maximize payoff $u_a$, $\mathcal{A}$ needs to assign 
  an amount of $p_{n-t+1}^{r}$ voting power to exactly $t$ (i.e., MinApprov) of $\mathcal{A}$'s candidates\footnote[8]{We assume that $\mathcal{A}$ wins in a tie vote.}, where $p_{n-t+1}^{r}$ stands for the voting power of $c_{n-t+1}^{r}$, the $(n-t+1)^{th}$ element in the vector $\mathbf{c^r}$.
  To illustrate this point, if $\mathcal{A}$ removes an amount of $\delta$ voting power from any of the $t$ candidates, the corresponding candidate will be defeated by $c_{n-t+1}^{r}$, resulting in $u_a=-\infty$. 
  In contrast, if $\mathcal{A}$ assigns an additional amount of $\delta$ voting power to any of $\mathcal{A}$'s candidates, $u_a$ will be decreased by $\delta$. 
  In either case, $u_a$ would become smaller than the payoff of $\mathcal{A}$ by taking the strategy $\widehat{s_a}$.
  % \textcolor{black}{This demonstrates that $\widehat{s_a}$ is indeed the unique best response in any of these subgames.}
\end{proof}

  % $\mathcal{A}$ is given a vector $\mathbf{c^r} = (c^r_1,...,c^r_n)$, which stands for $\mathcal{R}$'s candidates sorted by their voting power assigned by $\mathcal{R}$ from high to low.

\begin{lemma}
\label{lemma:2}
$\widehat{s_r}$ \textcolor{black}{is the best response} of $\mathcal{R}$ \textcolor{black}{to $\widehat{s_a}$}.
\end{lemma}

\begin{proof}
  \textcolor{black}{Given $\widehat{s_a}$}, to maximize $u_r$, $\mathcal{R}$ needs to assign an amount of $\frac{\zeta_r p_r}{n-t+1}$ voting power to exactly $(n-t+1)$ candidates in $C_l$,
  \textcolor{black}{which means that $p^r_{n-t+1} = \frac{\zeta_r p_r}{n-t+1}$.}
  To prove it, if $\mathcal{R}$ moves an amount of $\delta$ voting power from any of the $(n-t+1)$ candidates to another candidate in $C_l$, $p_{n-t+1}^{r}$ would be decreased by $\delta$, which decreases $u_r$ by $t\delta$. 
\end{proof}

% $\widehat{s_r}$ strictly dominates other strategies of $\mathcal{R}.$

% \begin{table}
%   \small
%   \begin{center}
%   \begin{tabular}{|p{0.8cm}|p{1.6cm}|p{1.5cm}|p{2.5cm}|}
%   \hline
%   {\textbf{Chain}} & {\textbf{$R_A$ (current)}} & {\textbf{$R_A$ (upper)}} & {\textbf{$R_A$ (upper + `lazy')}}  \\
%       \hline
%       EOSIO & $p_r$ & $2.14 p_r$ & $3 p_r$  \\
%       \hline % \hdashline
%       Steem & $p_r$ & $4.25 p_r$ & $5 p_r$  \\
%       \hline % \hdashline 
%       TRON  & $p_r$ & $2.11 p_r$ & $3 p_r$    \\
%       \hline
%   \end{tabular}
%   \end{center}
%   % \vspace{-2mm}
%   \caption{The current active resistance $R_A$ (left column) and the theoretical upper bound of $R_A$ by setting $v=n-t+1$.}
%   %  without (middle column) or with (right column) the `lazy' assumption.}  
%   \vspace{-8mm}    
%   \label{table:revisit}
% \end{table}

\noindent \textbf{The amplification effect.}
Let \textcolor{black}{us} now discuss $\zeta_a$ and $\zeta_r$, the two power amplification coefficients.
\textcolor{black}{
Intuitively, depending on whether a voting system allows voters to weight multiple votes using the same coins, an approval voting system tends to amplify voters' power by $v$ (i.e., MaxVote) while a cumulative voting system does not.
}
In a cumulative voting system, for both $\mathcal{A}$ and $\mathcal{R}$, every single unit of voting power can only be assigned to a single candidate and therefore, we can simply set both $\zeta_a$ and $\zeta_r$ as 1.
In an approval voting system, for both $\mathcal{A}$ and $\mathcal{R}$, every single unit of voting power can be assigned to up to $v$ distinct candidates and therefore, both $\zeta_a$ and $\zeta_r$ are upper-bounded by $v$.
However, the strategy $\widehat{s_a}$ suggests $\mathcal{A}$ to pick exactly $t$ candidates, which actually bounds $\zeta_a$ by $t$.
Similarly, $\widehat{s_r}$ bounds $\zeta_r$ by $n-t+1$.
To sum up, we have:
\begin{align}\zeta_a=\left\{\begin{aligned}
  min\{v,t\}\ (approval\ voting) \\ 
  1  \ \ \ \ \ \ \  \ \ \ \     (cumulative\ voting)
\end{aligned}\right.\label{e1}\end{align}
\begin{align}\zeta_r=\left\{\begin{aligned}
  min\{v,n-t+1\}\ (approval\ voting) \\
  1    \ \ \ \ \ \ \  \ \ \ \ \ \ \ \   \ \ \ \ \ \ \ \  (cumulative\ voting)
\end{aligned}\right.\label{e2}\end{align}

% {\bf Chao, the above discussion lacks the discussion of the intuition - there are mainly details that are less intuitive to quickly grasp. It will good to add some more points describing the intuition.}

\noindent \textcolor{black}{\textbf{The quantification.}}
Based on the above discussion, we can now \textcolor{black}{quantify the \textit{active takeover resistance}, $R_A$ introduced in Definition~\ref{def:active_resistance}, as the value of $p_a$ in the equilibrium}.
Next, in Lemma~\ref{lemma:3}, we quantify $R_A$ by combining both the two strategies $\widehat{s_a}$ and $\widehat{s_r}$ in the equilibrium.
Furthermore, we demonstrate the upper bound of $R_A$.

\begin{lemma}
\label{lemma:3}
\textcolor{black}{On the equilibrium path induced by $\widehat{s_r}$ and $\widehat{s_a}$ together,} the active takeover resistance $R_A = \frac{\zeta_r t p_r}{\zeta_a (n-t+1)}$, which is upper-bounded by $\frac{t p_r}{n-t+1}$ for a supermajority governance system where $\frac{2}{3}n<t<n$.
\end{lemma}

\begin{proof}
  Based on \textcolor{black}{Definition~\ref{def:active_resistance}}, Lemma~\ref{lemma:1} and Lemma~\ref{lemma:2}, we have $\zeta_a R_A = \textcolor{black}{\zeta_a p_a =\ } t p_{n-t+1}^{r} = t \frac{\zeta_r p_r}{n-t+1}$, so $R_A = \frac{\zeta_r t p_r}{\zeta_a (n-t+1)}$. 
  Next, based on Equation~\ref{e1} and Equation~\ref{e2}, given the cumulative voting rule, we have $\zeta_a=\zeta_r=1$, which makes $R_A=\frac{t p_r}{n-t+1}$.
  However, given the approval voting rule, $R_A$ is a piecewise function and is maximized when $v \le n-t+1$ and $\frac{2}{3}n<t<n$, which makes  $\zeta_a=\zeta_r=v$ and $R_A=\frac{t p_r}{n-t+1}$.
\end{proof}

Finally, based on the proof for Lemma~\ref{lemma:3}, we can easily prove Lemma~\ref{lemma:4}. 
% into $R_A = \frac{\zeta_r t p_r}{\zeta_a (n-t+1)}$.
% both $R_A = \frac{\zeta_r t p_r}{\zeta_a (n-t+1)}$ and $R_A=\frac{\lceil \frac{t}{v}  \rceil}{\lceil \frac{n-t+1}{v} \rceil} p_r$.

\begin{lemma}
  \label{lemma:4}
    Given a pair of parameters $(t,n)$ such that $\frac{2}{3}n<t<n$, by setting the MaxVote parameter $v \le n-t+1$, the active takeover resistance $R_A$ can achieve the upper bound, regardless of whether approval voting or cumulative voting is employed.
    %  whether or not the players are `lazy'.
  \end{lemma}

\begin{table}
  \small
  \begin{center}
  \begin{tabular}{|p{1cm}|p{2cm}|p{2cm}|}
  \hline
  {\textbf{Chain}} & {\textbf{$R_A$ (current)}} & {\textbf{$R_A$ (upper)}}  \\
      \hline
      EOSIO & $p_r$ & $2.14 p_r$  \\
      \hline % \hdashline
      Steem & $p_r$ & $4.25 p_r$  \\
      \hline % \hdashline 
      TRON  & $p_r$ & $2.11 p_r$  \\
      \hline
  \end{tabular}
  \end{center}
  % \vspace{-2mm}
  \caption{\small The current active resistance $R_A$ (left column) and the theoretical upper bound of $R_A$ by setting $v=n-t+1$.}
  %  without (middle column) or with (right column) the `lazy' assumption.}  
  \vspace{-8mm}    
  \label{table:revisit}
\end{table}

% Lemma~4 can be proved by 

% In a cumulative voting system, when both $\mathcal{A}$ and $\mathcal{R}$ are `lazy', 
% We omit the proof.

\begin{figure*}
  \centering
  %\vspace{-3 mm}
  \subfigure[{\small EOSIO}]
  {
     \label{vote_no_E}
      \includegraphics[width=0.64\columnwidth]{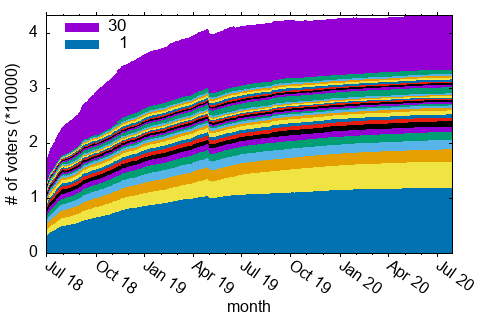}
  }
  \subfigure[{\small Steem}]
  {
    \label{vote_no_S}
      \includegraphics[width=0.64\columnwidth]{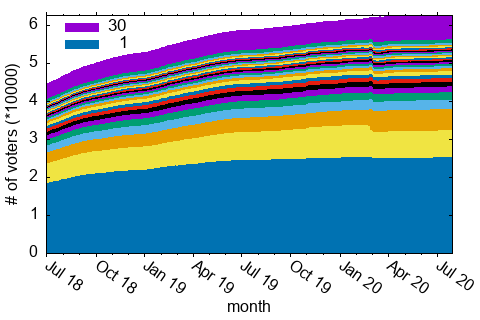}
  }
  \subfigure[{\small TRON}]
  {
    \label{vote_no_T}
      \includegraphics[width=0.64\columnwidth]{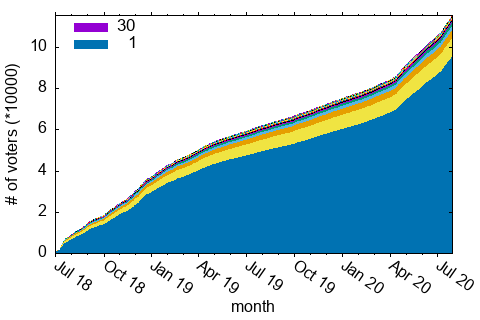}
  }
  \vspace{-4.5mm}
  \caption{\small The daily variations in the size of different voter categories, based on the number of votes cast.}
  \vspace{-3.0mm}
  \label{vote_no}
\end{figure*}

\begin{figure*}
  \centering
  %\vspace{-3 mm}
  \subfigure[{\small EOSIO}]
  {
     \label{priorities_E}
      \includegraphics[width=0.636\columnwidth]{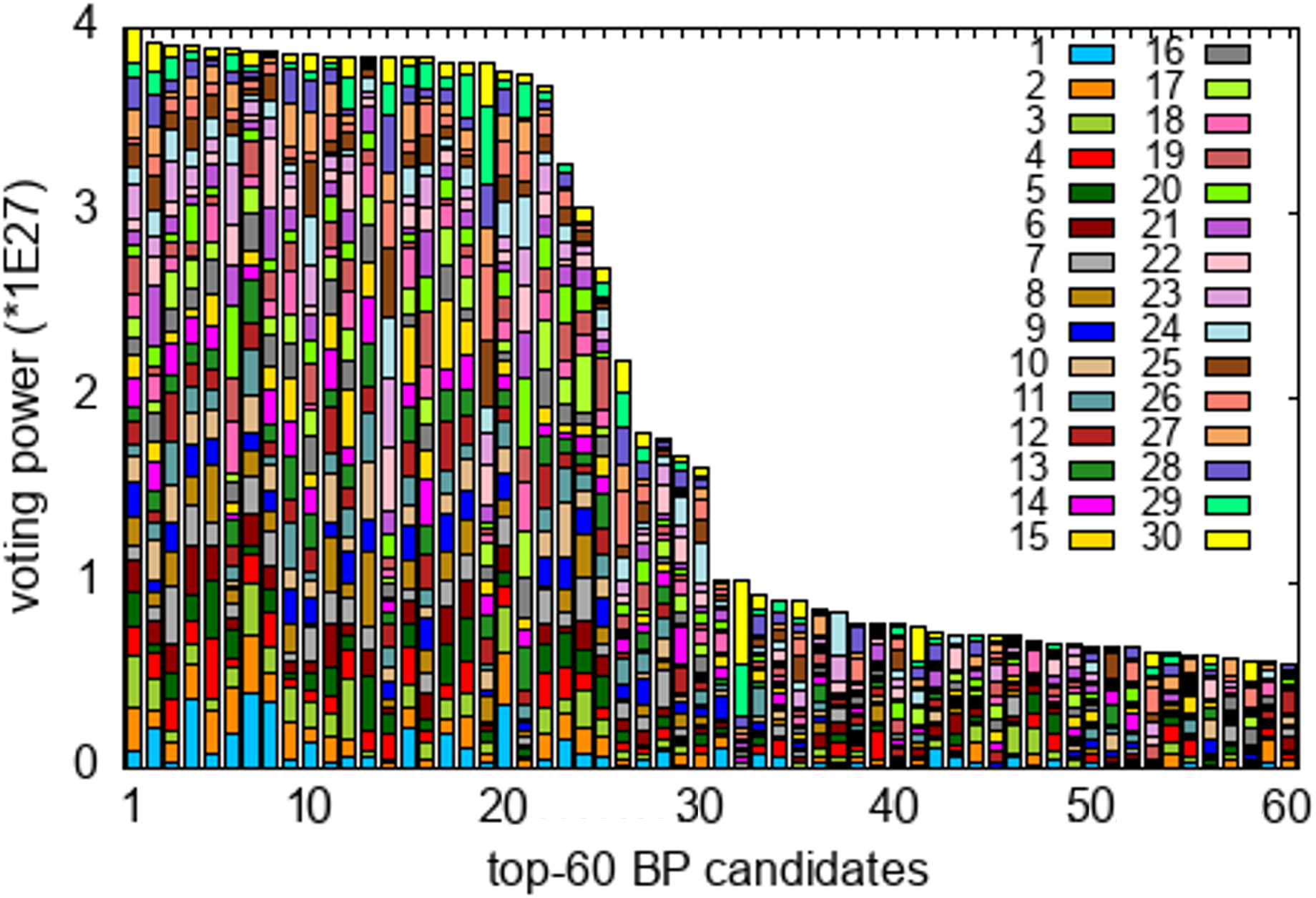}
  }
  \subfigure[{\small Steem}]
  {
    \label{priorities_S}
      \includegraphics[width=0.636\columnwidth]{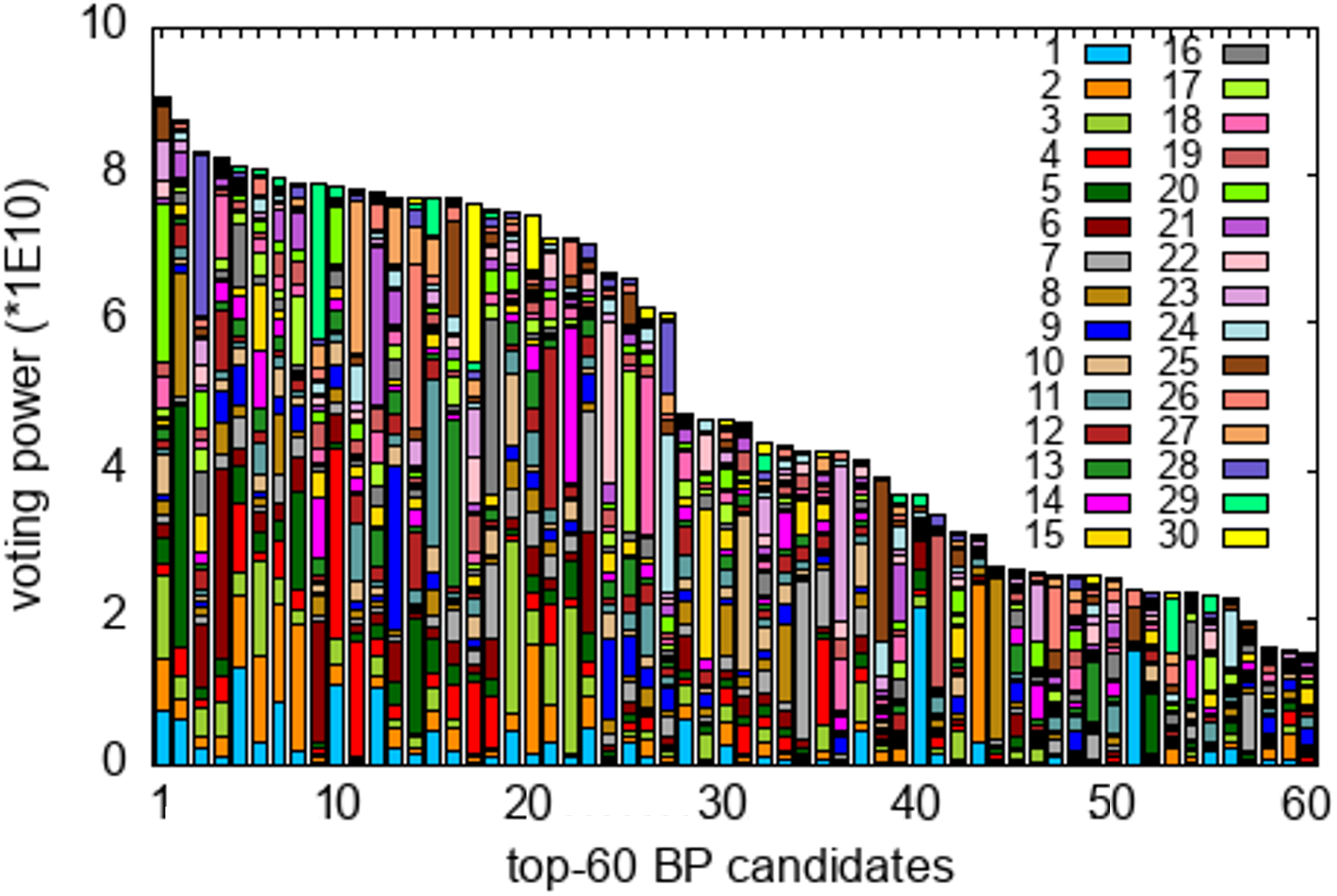}
  }
  \subfigure[{\small TRON}]
  {
    \label{priorities_T}
      \includegraphics[width=0.635\columnwidth]{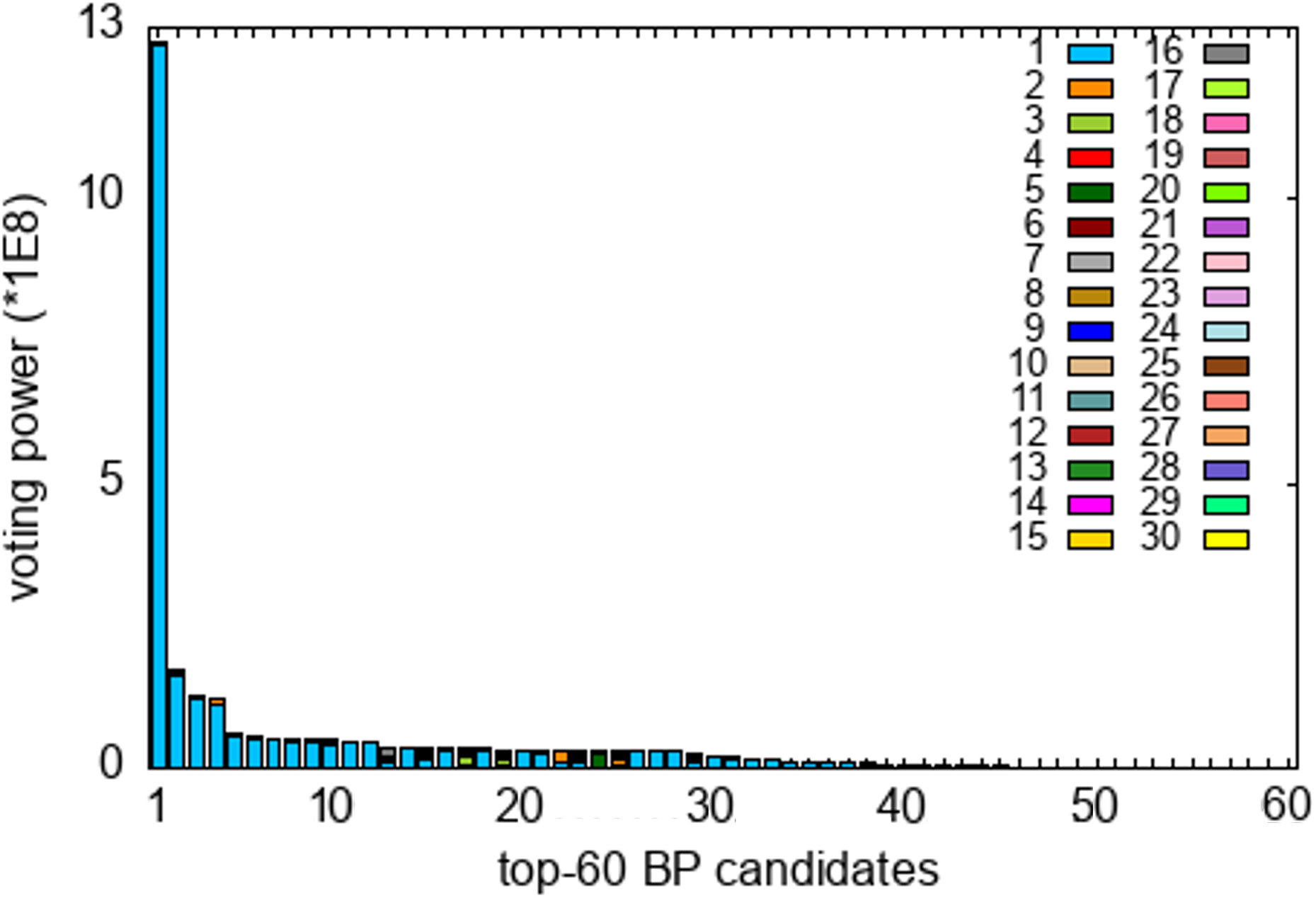}
  }
  \vspace{-4.5mm}
  \caption{\small The snapshot of the top-60 BP candidates on Feb. 14, 2020, where each candidate's voting power is divided into up to 30 segments based on the priorities assigned to them by voters.}
  \vspace{-2.0mm}
  \label{priorities}
\end{figure*}

\noindent \textbf{EOS, Steem and TRON.}
We now revisit\footnote[9]{For Steem, we ignore the rotational seat and set $n=20$.} the resistance $R_A$ deduced from the parameters of EOS, Steem and TRON shown in Table~\ref{table:summary}.
% {\bf For Steem, we ignore the rotational seat and set $n=20$.}
% {\bf Chao, I suggest moving the previous line as a footnote.}
The results shown in Table~\ref{table:revisit} demonstrate that the current resistance of DPoS blockchains is far below the theoretical upper bound.

\textcolor{black}{
While we now have the answer to our first research question, which is to maximize active resistance by setting $v \le n-t+1$, we discuss a more complex scenario, namely community-to-community takeover, in Appendix~\ref{appendix_B}.
Next, we answer our second research question via an empirical analysis.}

% {\bf Chao, in Table 2, it will be good to explicitly mention which column in the theoretical upper bound.}
%-------------------------------------------------------------------------------
\vspace{-0.6mm}
\section{Passive Resistance: an Empirical Analysis}
\label{sec6}
%-------------------------------------------------------------------------------
\textcolor{black}{In this section, we answer the second research question by performing the first large-scale empirical study of the passive takeover resistance of EOSIO, Steem and TRON.
We first describe the collected dataset and investigate the actual voter preferences, including the number of votes cast and the priorities assigned to chosen candidates.
Then, based on the observed voter preferences, we simulate the distribution of voting power under diverse voting system design choices and quantitatively evaluate the passive takeover resistance across different design choices using two metrics.
% We then simulate the  the phenomenon of voting power decay and measure the empirical resistance of EOSIO, Steem and TRON to takeovers based on the proposed metrics.
}
% We first describe the dataset that we collected and the metrics for evaluating the passive resistance to takeovers.
% We then investigate voter preferences and introduce the phenomenon of voting power decay.
%  and we measure the empirical resistance of EOSIO, Steem and TRON to takeovers based on the proposed metrics.}
% Finally, we discuss potential ways to improve the resistance of DPoS blockchains to takeovers.

\vspace{-0.6mm}
\subsection{Voter Preferences}
\label{s6.1}

\textcolor{black}{
We collected and parsed real data from the EOSIO, Steem, and TRON blockchains. Based on this dataset, we measure and analyze voters' number of votes cast and voting priorities.}

\iffalse
We have proved in Lemma~\ref{lemma:2} of Section~\ref{sec:game} that, in the equilibrium, the co-resisters will choose to assign an amount of $\frac{\zeta_r p_r}{n-t+1}$ voting power to exactly $(n-t+1)$ candidates.
However, voters in practice may not have followed this defensive strategy.
For instance, to defend Steem against TRON, the best strategy of voters is that every voter cast at least four votes and support the same four candidates, which was not followed by most voters in practice, as illustrated in Figure~\ref{takeover_snapshot_1}.
Therefore, to understand voters' actual voting preferences, based on our dataset, we measure and analyze two things:
(1) the number of votes that a voter was actually casting;
(2) the potential priorities that a voter actually assigned to her chosen candidates, namely the possible order of her votes to withdraw if the voter has to reduce the number of votes.
\fi

\noindent \textbf{Dataset.}
We collect the Steem blockchain data and TRON blockchain data using their official APIs~\cite{SteemAPI,TRONAPI} and obtain the EOSIO blockchain data from the dataset released by a recent work~\cite{zheng2021xblock}.
The basic information and statistics of our dataset are shown in Table~\ref{table:dataset}.
Based on this dataset, we construct per day power snapshots and also per day voting snapshots for all three blockchains.
Specifically, a power snapshot refers to a collection of \textit{<voter, voting power> } pairs by the end of a certain day, where a voter's voting power consists of her own voting power and voting power delegated to her, if any.
Similarly, a voting snapshot refers to a collection of \textit{<voter, candidates>} pairs by the end of a certain day. 
Based on these snapshots, we are capable of capturing daily changes in the blockchains to perform fine-grained empirical analysis.
Our empirical study presented in this section focuses on a period of two years from July 2018 to July 2020 so that we can compare the three blockchains after both EOS and TRON have been created in June 2018.
% Next, based on the dataset, we measure and analyze the number of votes cast by voters and their voting priority.

\noindent \textbf{No. of votes.}
We present the results of daily changes in the size of different voter categories, based on the number of votes cast (ranging from 1 to 30) as a stacked line chart in Figure~\ref{vote_no}. 
Surprisingly, even though voters in all three blockchains can cast up to 30 votes, many choose to cast only a few or, in some cases, a single vote. 
% The results of the daily change in the size of different categories of voters who were casting a different number of votes, ranging from 1 to 30, are shown as a stacked line chart in Figure~\ref{vote_no}.
% Surprisingly, we find that, even if voters are allowed to cast up to 30 voters in all three blockchains, many voters have still chosen to cast only a few votes and in some cases, even a single vote.
It may be easier to understand the phenomenon in TRON because TRON adopts the cumulative voting rule so that voters can not amplify their power by casting more votes.
Nevertheless,  we find that nearly half of EOSIO voters choose to cast fewer than 5 votes, and more than half of Steem voters consistently cast fewer than 3 votes.
There are many possible reasons that can drive voters to cast a few votes.
For instance, a voter may be recommended only a few candidates by a friend or an online article, may find it tedious to repeatedly click the vote buttons, and may belong to or be compromised or bribed by a single candidate.
The phenomenon may be desirable from perspectives such as diversity, but clearly not desirable from a perspective of protecting DPoS blockchains against takeovers. The fact that voters do not fully utilize the amplification effect potentially makes takeovers easier for attackers who understand and exploit the rule.
More concretely, the value of $(n-t+1)$ is 7 and 4 in current EOSIO and Steem, respectively.
However, over half of the voters in both blockchains cast fewer votes than the two equilibrium-suggested thresholds, implying that most voters may not consider takeover risks in practice.

\begin{table}
  \small
  \begin{center}
  \begin{tabular}{|p{1.5cm}|p{1.7cm}|p{1.7cm}|p{1.7cm}|}
  \hline
  {} & {\textbf{EOSIO}} & {\textbf{Steem}} & {\textbf{TRON}}  \\
      \hline
      Start date & 2016-03-24 & 2018-06-18 & 2018-06-25  \\
      \hline % \hdashline
      End date & 2020-07-31 & 2020-07-31 & 2020-07-31  \\
      \hline % \hdashline
      End block & 134,193,882 & 45,568,376 & 21,980,572  \\
      \hline % \hdashline 
      Voters & 56,119 & 67,605 & 115,508  \\
      \hline % \hdashline  
      Candidates  & 596 & 890 & 268   \\
      \hline
  \end{tabular}
  \end{center}
  % \vspace{-2mm}
  \caption{\small Basic information and statistics of the dataset.}    
  \vspace{-8.5mm}  
  \label{table:dataset}
\end{table}

\noindent \textbf{Voting priority.}
We have seen that voters often cast fewer votes than expected and it is actually quite common in practice.
Intuitively, DPoS blockchains should reduce the MaxVote parameter $v$ to minimize the gap between voters and attackers, due to their different preferences regarding the number of votes to cast.
However, it is then important to estimate the priorities that voters would assign to candidates.
For instance, in a voting system where $v=2$, voters A and B are voting for two sets of candidates (C,D) and (C,E), respectively, where candidate C is their shared choice.
Now, if we want to study the passive resistance of the system to takeovers in case of a smaller $v$ and thus reduce $v$ from 2 to 1, each voter will need to withdraw one vote and decide which one to remove.
The withdrawal order matters because if both voters A and B retain candidate C, the shared choice, after their withdrawal, their voting power will still be aggregated at candidate C. This potentially makes takeovers more difficult even if the aggregation occurs unintentionally.
We understand the difficulty in accurately estimating voters' behaviors due to the complexity of their motivations and the lack of ground truth data after altering system parameters.
Similar to recent works on approval voting~\cite{scheuerman2021modeling,scheuerman2019heuristics}, we propose a simple but reasonable heuristic and assume all voters follow it.
Specifically, we assume that a voter would assign the lowest priority to the newly selected candidate while the highest priority to the candidate that the voter has voted for the longest time.
In other words, we regard the candidates chosen by a voter as a vector, sorted by the duration they have remained in the vector, and assume that voters will remove the last candidate from the vector first.

In Figure~\ref{priorities}, we present a stacked bar chart illustrating the snapshot of the top 60 BP candidates  as of Feb. 14, 2020, which is half a month before TRON's takeover. Each candidate's voting power is divided into up to 30 segments, with segment $i$ representing the voting power contributed by voters who assigned the $i^{th}$ priority to the candidate.
The results reveal several interesting characteristics.
From a macro perspective, we observe that the voting power in EOSIO is more concentrated among the top 22 candidates and declines rapidly beyond the $31^{st}$ candidate.
In contrast, voting power in Steem decreases more smoothly.
In TRON, however, we find that the first BP receives an overwhelming amount of voting power, over 7 times that of the second BP.
% We will discuss the implications of this phenomenon on takeovers later in this section.
From a micro perspective, we note that in EOSIO, voters tended to be highly inconsistent with the priorities assigned to candidates, indicating that voters do not generally assign their top-$k$ priorities to the same candidates.
Again, the relatively even distribution of priorities may not be desirable for resisting takeovers, as it suggests that voting power may not become more concentrated when the MaxVote parameter $v$ is reduced.
In contrast, priorities assigned to the top 12 candidates in TRON are dominated by the first priority, which is not surprising given the large proportion of voters casting a single vote.
% However, we also find that the domination of the first priority in TRON tended to become lower from the $13^{th}$ candidate.

\textcolor{black}{
In summary, we observe that EOSIO voters exhibit the most diverse preferences from a micro perspective, whereas TRON voters demonstrate the highest consistency in their preferences.}

\iffalse
\begin{figure*}
  \centering
  %\vspace{-3 mm}
  \subfigure[{\small EOSIO}]
  {
     \label{decay_E}
      \includegraphics[width=0.66\columnwidth]{./figures/decay_E.png}
  }
  \subfigure[{\small Steem}]
  {
    \label{decay_S}
      \includegraphics[width=0.66\columnwidth]{./figures/decay_S.png}
  }
  \subfigure[{\small TRON}]
  {
    \label{decay_T}
      \includegraphics[width=0.66\columnwidth]{./figures/decay_T.png}
  }
  \vspace{-2mm}
  \caption{The voting power decay phenomenon by reducing the MaxVote parameter $v$ from 30 to 1, where $v=x$ refers to an approval voting system with $v=x$ and $CV$ refers to a cumulative voting system with $v=30$.}
  \vspace{-2mm}
  \label{decay}
\end{figure*}
\fi

\vspace{-0.9mm}
\subsection{Passive Takeover Resistance}
\label{sec6.2}
\textcolor{black}{Next, based on the actual dataset and voter preferences, we simulate the voting power distribution for EOSIO, Steem, and TRON when adopting different voting system design choices. We then quantitatively evaluate the passive takeover resistance of these blockchains under various voting system design choices, using two metrics.}

\begin{figure*}
  \centering
  %\vspace{-3 mm}
  \subfigure[{\small EOSIO}]
  {
    %  \label{metric_1_E}
      \includegraphics[width=0.64\columnwidth]{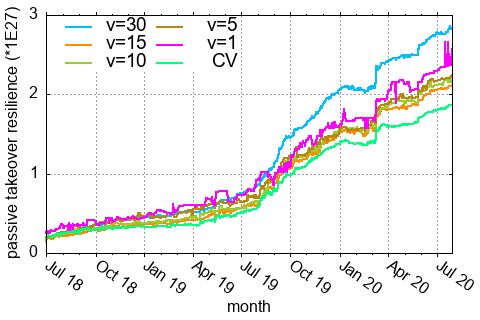}
  }
  \subfigure[{\small Steem}]
  {
    % \label{metric_1_S}
      \includegraphics[width=0.64\columnwidth]{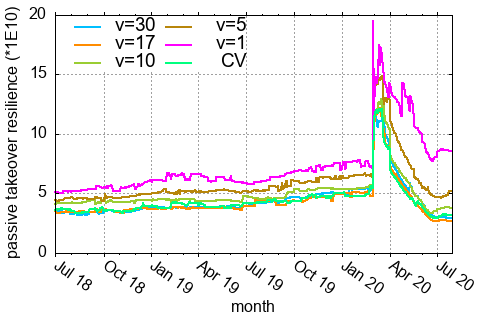}
  }
  \subfigure[{\small TRON}]
  {
    % \label{metric_1_T}
      \includegraphics[width=0.64\columnwidth]{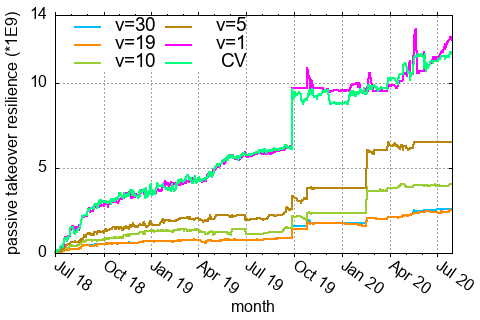}
  }
  \vspace{-5mm}
   \caption{\small The daily variations of $R_P$, the \textit{passive takeover resistance}.}
  \vspace{-3.5mm}
  \label{metric_1}
\end{figure*}

\begin{figure*}
  \centering
  %\vspace{-3 mm}
  \subfigure[{\small EOSIO}]
  {
    %  \label{metric_1_E}
      \includegraphics[width=0.64\columnwidth]{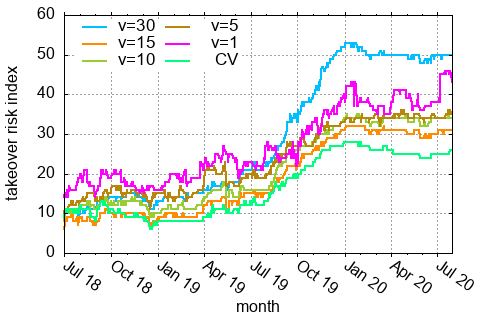}
  }
  \subfigure[{\small Steem}]
  {
    % \label{metric_1_S}
      \includegraphics[width=0.64\columnwidth]{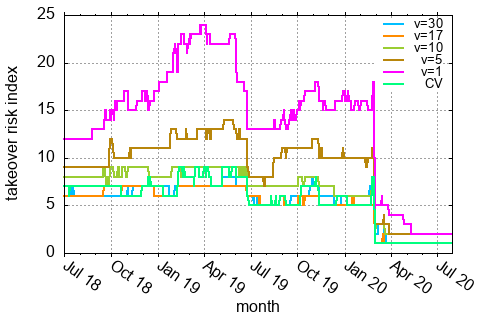}
  }
  \subfigure[{\small TRON}]
  {
    % \label{metric_1_T}
      \includegraphics[width=0.64\columnwidth]{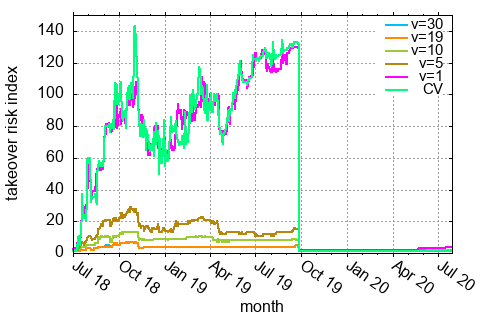}
  }
  \vspace{-5.0mm}
  \caption{\small The daily variations of $I_t$, the \textit{takeover risk index}.}
  \vspace{-2.0mm}
  \label{metric_2}
\end{figure*}

% \subsection{Voting Power Decay}
\noindent \textbf{\textcolor{black}{Simulation}.}
For EOSIO, Steem, and TRON, we simulate scenarios where the blockchain employs an approval voting system with a fixed pair of $(t,n)$, as displayed in Table~\ref{table:summary}, and a MaxVote $v$ varying from 30 to 1.
Additionally, we simulate situations where the blockchain utilizes a cumulative voting system with $v=30$, as in TRON.
As previously observed in Section~\ref{s6.1}, the diversity in voter preferences naturally leads to a phenomenon we term \textit{voting power decay}, which refers to the decrease in voting power capable of passively resisting takeovers as the MaxVote parameter $v$ is reduced.
However, voter behavior may exhibit a certain level of uncertainty after modifying the voting system.
Consequently, we made several assumptions during the simulation process.
Specifically, to simulate an approval voting system, we assumed that voters would withdraw their votes based on their priorities as $v$ is reduced.
Moreover, we assumed that a voter in TRON, upon adopting the approval voting rule, would give all her votes the weight of her full voting power.
To simulate a cumulative voting system, we assumed that voters who cast multiple votes in EOSIO and Steem would evenly distribute their voting power among all the candidates they select, which is the most commonly observed heuristic in TRON.
% To gain a comprehensive understanding of the phenomenon, for each blockchain, we simulate the situations that the blockchain employs an approval voting system with a fixed pair of $(t,n)$ as displayed in Table~\ref{table:summary} and a MaxVote $v$ that varies from 30 to 1, as well as the situations that the blockchain employs a cumulative voting system where $v=30$ as in TRON.

\iffalse
In Figure~\ref{decay}, by reducing MaxVote $v$ from 30 to 1, we measure the decay of the voting power of the $(n-t+1)^{th}$ candidate, namely $p_{n-t+1}$ in Equation~\ref{e3}, in different settings of the voting system.
In both EOSIO and Steem, we observe a clear layered structure of the voting power decay while the layers in TRON tend to be less structured because the $(n-t+1)^{th}$ candidate in TRON may sometimes receive no voting power that corresponds to a certain priority.
Besides, by observing temporal variation of $p_{n-t+1}$, we could see that, in EOSIO, $p_{n-t+1}$ has started to grow faster since Aug. 19.
In Steem, $p_{n-t+1}$ has changed little until TRON's takeover on Mar, 20, where a surge can be clearly observed.
In TRON, $p_{n-t+1}$ has tended to be relatively stable until Oct. 19, during which a single voter converted a significant amount of coins into voting power.
\fi

\noindent \textbf{Metrics.}
We propose two metrics to evaluate the passive resistance to takeover attacks.
\textcolor{black}{The first metric quantifies the \textit{passive takeover resistance}, $R_P$, in the way described in Definition~\ref{def:passive_resistance}.
Specifically, $R_P$ corresponds to the \textit{active takeover resistance} $R_A$ and measures the minimum amount of voting power that an attacker needs to have to take over a blockchain.}
Recall that for active resistance, as illustrated in Lemma~\ref{lemma:3}, we have $\zeta_a R_A = t p_{n-t+1}^{r}$ in equilibrium. 
However, for passive resistance, instead of assuming that the $(n-t+1)^{th}$ BP is controlled by the co-resisters, we disregard the BP's attitude towards takeovers and directly measure the actual voting power of the $(n-t+1)^{th}$ BP, denoted as $p_{n-t+1}$.
Therefore, in a voting system with parameters $(v,t,n)$, $R_P$ can be computed as $\zeta_a R_P = t p_{n-t+1}$, namely $R_P = \frac{t p_{n-t+1}}{\zeta_a}$.
% \begin{align}
%   \label{e3}
%     R_P = \frac{t p_{n-t+1}}{\zeta_a} 
% \end{align}
% \noindent where $p_{n-t+1}$ represents the voting power of the $(n-t+1)^{th}$ BP after sorting all BPs based on their voting power.

\iffalse
is the minimum amount of voting power that an attacker needs to have to successfully take over a blockchain, which is used to quantify the \textit{passive takeover resistance} and denoted as $R_P$.
In a voting system with parameters $(v,t,n)$, $R_P$ can be computed as:
% $$R_P = \frac{t p_{n-t+1}}{\zeta_a} $$
\begin{align}
\label{e3}
  R_P = \frac{t p_{n-t+1}}{\zeta_a} 
\end{align}
\noindent where $p_{n-t+1}$ represents the voting power of the $(n-t+1)^{th}$ BP after sorting all BPs based on their voting power.
Recall that, as illustrated in Lemma~\ref{lemma:3}, in the equilibrium, we have $\zeta_a R_A = t p_{n-t+1}^{r}$. In the empirical analysis, however, instead of assuming that the $(n-t+1)^{th}$ BP is controlled by the co-resisters, we directly measure the actual voting power.
\fi

The second metric, referred to as the \textit{takeover risk index}, is denoted as $I_t$.  
It measures the minimum number of voters whose combined voting power can successfully take over a blockchain.
To compute $I_t$ for a given day, we first sort voters based on the power snapshot of that day and obtain a vector of sorted voters. We then determine the minimum value of $i$ as $I_t$ such that the sum of voting power owned by the top-$i$ voters in the vector exceeds $R_P$.
Based on $R_P$ and power snapshots, the \textit{takeover risk index} captures the risks associated with top voters who, although individually incapable of taking over a blockchain, may collude to do so.
A metric with a similar purpose, called the Nakamoto coefficient~\cite{fritsch2022analyzing,lin2021measuring}, has been widely used to evaluate blockchain decentralization. It essentially measures the minimum number of resource holders (e.g., mining power) required for their combined resources to surpass a threshold (e.g., 50\%).
However, we could not directly employ the Nakamoto coefficient in our study because each voter in a DPoS blockchain may vote for a different set of BP candidates while only the votes received by the top $(n-t+1)$ BPs may affect the difficulty of a takeover attack. As a result, we propose the \textit{takeover risk index}.

% \balance

% \subsection{Voting Preferences}

%  and will keep retaining the first candidate as long as the voter is allowed to cast a single vote.

% \subsubsection{Evaluation}
\noindent \textbf{Evaluation.}
% After obtaining $p_{n-t+1}$, we can now start computing the two metrics presented in Section~\ref{sec:dataset}, namely $p_a^m$ the \textit{empirical takeover resistance} and $I_t$ the \textit{takeover risk index}.
% Concretely, by following Equation~\ref{e3}, we can multiply $p_{n-t+1}$, which has been presented in Figure~\ref{decay}, by $\frac{t}{\zeta_a}$ where the value of $\zeta_a$ can be computed by inserting the type of voting system and the value of $(v,t)$ to Equation~\ref{e1}.
We start by evaluating the \textit{passive takeover resistance}, $R_P = \frac{t p_{n-t+1}}{\zeta_a}$, for EOSIO, Steem, and TRON under a range of voting system design choices.
Theoretically, based on Equation~\ref{e1}, we can categorize these design choices into three groups:
% (1) Group 1: For an approval voting (AV) system with $1<v \le t$, we have $\zeta_a=v$ and thus $R_P = \frac{t}{v} p_{n-t+1} \propto \frac{p_{n-t+1}}{v}$;
% (2) Group 2: For an approval voting (AV) system with $t<v \le 30$, we have $\zeta_a=t$ and thus $R_P = p_{n-t+1}$;
% (3) Group 3: For a cumulative voting (CV) system, $\zeta_a=1$, so $R_P = t p_{n-t+1} \propto p_{n-t+1}$.
% \end{itemize}
\vspace{-0.5mm}
\begin{itemize}[leftmargin=*]
  \item Group 1: For an approval voting (AV) system with $1<v \le t$, we have $\zeta_a=v$ and thus $R_P = \frac{t}{v} p_{n-t+1} \propto \frac{p_{n-t+1}}{v}$.
  \item Group 2: For an approval voting (AV) system with $t<v \le 30$, we have $\zeta_a=t$ and thus $R_P = p_{n-t+1}$.
  \item Group 3: For a cumulative voting (CV) system, $\zeta_a=1$, so $R_P = t p_{n-t+1} \propto p_{n-t+1}$.
\end{itemize}
\vspace{-0.5mm}
% First, for an approval voting system with $1<v \le t$, we have $\zeta_a=v$ and thus $R_P = \frac{t}{v} p_{n-t+1} \propto \frac{p_{n-t+1}}{v}$.
% Next, for an approval voting system with $t<v \le 30$, we have $\zeta_a=t$ and thus $R_P = p_{n-t+1}$.
% Finally, for a cumulative voting system, $\zeta_a=1$, so $R_P = t p_{n-t+1} \propto p_{n-t+1}$.
For design choices in the first group, we aim to understand how resistance $R_P$ changes when both the reduction of the $(n-t+1)^{th}$ BP's actual voting power $p_{n-t+1}$ (i.e., voting power decay, detrimental to resistance) and the reduction of the MaxVote parameter $v$ (i.e., weakened amplification effect, beneficial to resistance) occur. 
In the second group, we can anticipate that resistance $R_P$ will increase when the MaxVote parameter $v$ is larger. However, we want to determine if the design choice of employing a large $t$, say $t=30$, is the optimal option among the three groups. Finally, we aim to compare the two types of voting systems.

The daily variations in \textit{passive takeover resistance} $R_P$ are depicted in Figure~\ref{metric_1}.
Using the notation of $(AV, i)$ for an approval voting system with a MaxVote $v = i$ and $CV$ for a cumulative voting system, we can rank the design choices according to their respective resistance $R_P$, from best to worst, using the symbol `$\ge$' to indicate `slightly better,' and `$\approx$' to signify `almost the same':

\iffalse
\vspace{1mm}
\begin{mdframed}[innerleftmargin=1.6pt]
  \begin{packed_enum}[leftmargin=*]
    \begin{itemize}[leftmargin=*]
      \item $EOSIO: (AV,30)>(AV,1)\ge\cdots\ge(AV,t)>CV$ 
      \item $Steem: (AV,1)>(AV,2)>\cdots>CV>(AV,t)$
      \item $TRON: CV \approx (AV,1)>\cdots>(AV,t)\approx(AV,30)$
    \end{itemize}
  \end{packed_enum}
\end{mdframed}
\vspace{1mm}
\fi

\vspace{0.5mm}
\noindent\fbox{%
\parbox{\dimexpr\linewidth-2\fboxsep-2\fboxrule}{
  \begin{enumerate}[leftmargin=*,label={}]

    \vspace{-3mm}
\item
\begin{equation*}
  EOSIO: \underset{\text{group 2}}{\underbracket[0.5pt]{(AV,30)}}> 
  \underset{\text{group 1}}{\underbracket[0.5pt]{(AV,1)\ge\cdots\ge(AV,t)}}>
  \underset{\text{group 3}}{\underbracket[0.5pt]{CV}}
\end{equation*}
\vspace{-1mm}

\vspace{-4mm}
\item
\begin{equation*}
  Steem: \underset{\text{group 1}}{\underbracket[0.5pt]{(AV,1)>\cdots>(AV,t)}}\approx 
  \underset{\text{group 2}}{\underbracket[0.5pt]{(AV,30)}} \approx 
  \underset{\text{group 3}}{\underbracket[0.5pt]{CV}} 
\end{equation*}
\vspace{-1mm}

\vspace{-4mm}
\item
\begin{equation*}
  TRON: \underset{\text{group 3}}{\underbracket[0.5pt]{CV}} \approx 
  \underset{\text{group 1}}{\underbracket[0.5pt]{(AV,1)>\cdots>(AV,t)}}\approx
  \underset{\text{group 2}}{\underbracket[0.5pt]{(AV,30)}}
\end{equation*}
\vspace{-3mm}
\end{enumerate}
}}
\vspace{0.5mm}

% The results of the daily changes of $R_P $ are shown in Figure~\ref{metric_1}.
% By expressing an approval voting system with $v=i$ as $(AV,i)$ and a cumulative voting system as $CV$, we can rank the design choices based on their corresponding resistance $R_P$ from best to worst as follow, where `$\ge$' denotes `slightly better' and `$\approx$' denotes `almost same':
\iffalse
\begin{itemize}[leftmargin=*]
  \item $EOSIO: (AV,30)>(AV,1)\ge\cdots\ge(AV,t)>CV$ 
  \item $Steem: (AV,1)>(AV,2)>\cdots>CV>(AV,t)$
  \item $TRON: CV \approx (AV,1)>\cdots>(AV,t)\approx(AV,30)$
\end{itemize}
\fi
% $EOSIO: (AV,30)>(AV,1)\ge\cdots\ge(AV,t)>CV$ \\
% $Steem: (AV,1)>(AV,2)>\cdots>CV>(AV,t)$ \\
% $TRON: (AV,1)\approx CV>\cdots>(AV,t)\approx(AV,30)$

% \begin{align}
%   EOSIO: (AV,30)>(AV,1)\ge\cdots\ge(AV,t)>CV
% \end{align}
% \begin{align}
%   Steem: (AV,1)>(AV,2)>\cdots>CV>(AV,t)
% \end{align}
% \begin{align}
%   TRON: (AV,1)\approx CV>\cdots>(AV,t)\approx(AV,30)
% \end{align}

\noindent 
\textcolor{black}{
The results effectively address our questions.
For design choices in the first group, namely $(AV,1)$ to $(AV,t)$, we can see that the resistance $R_P$ tends to be higher for a smaller $v$ across all three blockchains. This implies that the impact of weakened amplification effect outweighs that of voting power decay.
However, since EOSIO voters exhibit the most diverse preferences from a micro perspective, the voting power decay in EOSIO is the most pronounced, leading to all design choices in the first group offering relatively similar levels of resistance for EOSIO.
For the second group, we note that $(AV,30)$ is the best choice of EOSIO but the worst choice of both Steem and TRON. This suggests that an approval voting system with a large $v$ might be more suitable for blockchains with more diverse voter preferences.
In contrast, we find that $CV$ is generally the worst choice for both EOSIO and Steem, while being the best choice for TRON. This indicates that $CV$ may be more fitting for blockchains where the voter preferences are highly consistent.}

Finally, we present the daily variations of the \textit{takeover risk index} $I_t$ in Figure~\ref{metric_2}.
Recall that $I_t$ is related to both $R_P$ and the power snapshots, resulting in similar trends over time for $I_t$ and $R_P$, as demonstrated by comparing Figure~\ref{metric_2} with Figure~\ref{metric_1}.
It is evident that EOSIO generally exhibits a larger $I_t$ than Steem, due to the more skewed distribution of voting power in Steem.
During the month of TRON's takeover, we note that $I_t$ of Steem drops to 1, which highlights  the capability of $I_t$ to detect known events.
More interestingly, we find that $I_t$ of TRON reaches 1 in Oct. 2019, indicating the presence of a single voter with sufficient voting power to take over TRON, which demonstrates the effectiveness of $I_t$ in identifying unknown takeover risks.

% {\bf the analysis of the results is very detailed and thorough. It will be good if we can provide some form of a road map for what we aim to see in the results and its significance and then we can just analyze the result and show that we observe what we aimed to see. In the present form, the discussion is heavy on the analysis and less text is on the road map and significance.}

\textcolor{black}{
In summary, our findings indicate that the approval voting rule with a small MaxVote parameter $v$ is a suitable choice for all three blockchains examined in this work.
% , and this choice may be generalizable to additional blockchains.
The consistency of our findings across multiple blockchains demonstrates the robustness of our conclusions and implies that our recommendations may serve as a foundation for optimizing voting system design choices in diverse DPoS blockchain environments.
Furthermore, for blockchains with more diverse voter preferences, such as EOSIO, the approval voting rule with a large $v$ may also help improve passive resistance.
Conversely, for blockchains with less diverse voter preferences, like TRON, the cumulative voting rule may be a viable alternative.
}

\vspace{-1mm}
\section{Discussion}
\label{sec:discussion}
% \noindent \textbf{Insights based on our empirical analysis.}
% We have seen that the approval voting rule with a small MaxVote parameter $v$ is a good choice for all the three blockchains that we have investigated in this paper and this choice may be generalized to more blockchains.
% Also, for blockchains where voters' voting preferences are more diverse, such as EOSIO, the approval voting rule with a large $v$ may also help improve resistance.
% For blockchains where voters' voting preferences are less diverse, such as TRON, the cumulative voting rule may also be a good choice.

\noindent \textbf{A hybrid approach.}
As illustrated by our measurements of the \textit{takeover risk index}, 
the passive resistance alone may be inadequate to resist takeovers as the voters may not be capable of understanding and taking advantage of the voting rule.
Besides the selection of the most appropriate voting system design choices,
the community of a blockchain may also arrange an amount of dedicated voting power, which is only used for actively preventing takeovers without affecting the election.
Specifically, the dedicated voting power can be delegated or transferred to a smart contract or a trusted party, which will continuously rank BP candidates based on the distribution of voting power excluding the dedicated part and leveraging the dedicated voting power to vote for exactly the top $(n-t+1)$ candidates.
\textcolor{black}{
For instance, in Steem, by setting $v=n-t+1=4$ and assigning an amount of dedicated voting power, $p_r$ to the top-4 candidates, the overall takeover resistance would become the sum of the passive takeover resistance $R_P$ and the upper-bounded active takeover resistance $R_A=4.25p_r$.
We believe that a hybrid approach that combines both passive and active resistance may provide a promising solution to improve takeover resistance.}

\begin{table}
  \small
  \begin{center}
  \begin{tabular}{|p{3.9cm}|p{3.8cm}|}
  \hline
  {\textbf{Consensus protocols}} & {\textbf{Blockchains}}   \\
      \hline
       \textbf{DPoS+PoA:} \newline  Proof of Staked Authority (PoSA), HPoS  & Binance Coin (BNB, \#4), Huobi Token (HT, \#56), KuCoin Token (KCS, \#57)  \\
      \hline % \hdashline
      \textbf{DPoS+BFT:} \newline Tendermint, Delegated Byzantine Fault Tolerance (dBFT) & Cosmos (ATOM, \#20),  OKB (OKB, \#29), Terra Classic (LUNA, \#44), Neo (NEO, \#72),  Osmosis (OSMO, \#80), Kava (KAVA, \#98) \\
      \hline % \hdashline 
      \textbf{Liquid} Proof of Stake (LPoS)  & Tezos (XTZ, \#48)   \\
      \hline % \hdashline
      \textbf{Nominated} Proof of Stake (NPoS) & Polkadot (DOT, \#12)  \\
      % \hline % \hdashline
      % Delegated Byzantine Fault Tolerance (dBFT) & Neo (NEO, \#72)  \\
      \hline % \hdashline
      XinFin DPoS (XDPoS) & XDC Network (XDC, \#95)  \\
      % \hline % \hdashline
      % Polygon PoS & Polygon (MATIC, \#10)  \\
      \hline
  \end{tabular}
  \end{center}
  % \vspace{-2mm}
  \caption{\small \textcolor{black}{Recent variants of DPoS that implement coin-based voting governance and their associated Top 100 cryptocurrencies on coinmarketcap.com as of Jan. 15, 2023~\cite{coinm}.}}
  %  without (middle column) or with (right column) the `lazy' assumption.}  
  \vspace{-6mm}    
  \label{table:generalization}
\end{table}

\noindent \textbf{Generalization.}
\textcolor{black}{
In general, our analysis in this paper is applicable to any blockchain that employs the coin-based voting governance model introduced in Section~\ref{sec_model_governance}, including but not limited to the ones listed in Table~\ref{table:generalization}.
On the one hand, these blockchains inherit the core coin-based voting governance model from DPoS, making them vulnerable to takeover attacks.
On the other hand, they have made improvements upon the original DPoS~\cite{larimer2014delegated}, either by combining DPoS with other consensus protocols (e.g., PoA~\cite{ekparinyaattack}, BFT~\cite{buchman2016tendermint}) or by refining specific steps in the original DPoS (e.g., the committee size is dynamically adjustable in LPoS~\cite{allombert2019introduction}, BP candidates require nomination by others in NPoS~\cite{wood2016polkadot}). 
We believe that the work presented in this paper will lay out the foundation for enhancing the takeover resistance of these blockchains and can provide valuable insights for future research on the impact of new features on takeover resistance in potential new variants of DPoS.
}

\noindent \textbf{Limitation and future work.}
\textcolor{black}{
In Section~\ref{sec6.2}, we adopt certain assumptions for the sake of simplicity and manageability in our simulation. These assumptions are based on empirical data and applied uniformly across all voters. 
However, we recognize the potential for more accuracy in future studies by diversifying these assumptions, such as classifying voters into strategic and non-strategic categories.
Besides, in terms of future research directions, we have identified various recent DPoS variants that implement coin-based voting governance in Table~\ref{table:generalization}. A careful evaluation of the enhancements these variants bring to the original DPoS could provide valuable insights into their alignment or divergence with the governance model we propose. This method helps to understand the direct applicability of our models and findings and potentially reveal new research problems. For instance, our insights could be directly applied to DPoS+PoA blockchains as their governance model remains unaffected. In contrast, applying our findings to LPoS blockchains, which have an adjustable committee size, might require adjustments and hence present a new research problem.
}

\section{Related work}
\label{sec:related_work}
%-------------------------------------------------------------------------------
\noindent \textbf{Decentralization on blockchains.}
As the most prominent PoW blockchains, Bitcoin and Ethereum's decentralization have attracted sustained interest from researchers. 
% Currently, researchers generally agree that Bitcoin is already highly centralized.
In 2014, Gervais et al. conducted an empirical study of Bitcoin data in~\cite{gervais2014bitcoin}, and their results showed that many key processes in Bitcoin are substantially controlled by a few entities.
Subsequently, Feld et al. analyzed the peer-to-peer network of Bitcoin and focused in~\cite{feld2014analyzing} and concluded that the network is highly centralized.
% Their results showed that the distribution of the Bitcoin peer-to-peer network is concentrated in a few autonomous domains, reflecting a high degree of centralization.
Miller et al. further investigated the topology of the Bitcoin network in~\cite{miller2015discovering} and found that a small number of top 2\% nodes essentially controlled about 75\% of the effective resources.
% They argued that this phenomenon may run counter to the ideal vision, where resources are held in a balanced manner by nodes.
Zeng et al. measured the decentralization in Ethereum at the level of mining pool participants in~\cite{zeng2021characterizing}.  
% The results indicated that decentralization measured at a deeper level could be quite different from that measured across mining pools.
% With the success of Bitcoin, new PoW blockchain systems continue to emerge. 
After that, researchers have conducted a comparative analysis of the decentralization of different PoW blockchains.
In 2018, Gencer et al. compared the actual degree of decentralization of Bitcoin and Ethereum in~\cite{gencer2018decentralization}.
The results showed that Bitcoin and Ether were similarly decentralized.
% and there was no evidence that one was significantly more decentralized than the other.
In 2019, Kwon et al. studied the gaming of Bitcoin and Bitcoin Cash 
% (hard forked from Bitcoin in 2017) 
in~\cite{kwon2019bitcoin}. 
% The two systems are inherently homogeneous, resulting in miners being able to switch systems to mine at will, and the miner's position is thus fickle. 
This work modeled the mining game of these two systems and demonstrated that the Nash equilibrium of the game leads to severe centralization of the disadvantaged system. 
% With the rapid development of core blockchain technologies such as consensus protocols, a large number of blockchain systems based on non-PoW (e.g. PoS, DPoS) consensus protocols have gained a lot of attention, however, the research on non-PoW blockchain centralization is still in the early stage.
Recently, blockchains based on non-PoW consensus protocols have gained a lot of attention.
Kwon et al. analyzed the decentralization in various blockchains including PoW, PoS and DPoS in~\cite{kwon2019impossibility}.
Li et al. compared the decentralization between Steem and Bitcoin in~\cite{li2020comparison}.
% However, without fixing the missing system parameter $\lambda$, these studies only measured a snapshot of decentralization in Steem.
% However, without fixing the missing sys

\noindent \textbf{Attacks on blockchains.}
In 2014, Eyal et al. questioned whether Bitcoin incentives can achieve incentive compatibility in~\cite{eyal2014majority}. Their paper proposes a selfish mining attack, in which a selfish mining pool does not disclose new blocks mined to maintain its advantage, but discloses new blocks mined when it is about to lose its advantage. 
% The paper shows that a pool holding more than 33\% of the total resources could earn more rewards through selfish mining. 
Since then, Sapirshtein et al. optimized the selfish mining attack method in~\cite{sapirshtein2016optimal} and proposed an algorithm to define a lower bound on the resources an attacker needs to hold to benefit from selfish mining.
% Feng et al. studied selfish mining in Ethereum in~\cite{feng2019selfish} and they showed that the percentage of resources required to benefit from selfish mining in Ethereum tends to be is lower.
Gervais et al. proposed a quantitative framework that helps devise optimal adversarial strategies for double-spending and selfish mining in existing PoW-based deployments and PoW blockchain variants~\cite{gervais2016security}, which inspired our efforts to enhance takeover resistance in DPoS blockchains and their variants.

% The paper presents a quantitative framework for analyzing security and performance implications of PoW blockchains with different consensus and network parameters. It helps devise optimal adversarial strategies for double-spending and selfish mining, allowing for objective comparisons of tradeoffs between performance and security in various PoW deployments.

% Since then, Tschorsch et al. optimized the selfish mining attack method in~\cite{sapirshtein2016optimal} and proposed an algorithm to define a lower bound on the resources an attacker needs to hold to benefit from selfish mining.
% Feng et al. studied selfish mining in Ethereum in~\cite{feng2019selfish} and their results showed that the percentage of resources required to benefit from selfish mining in Ethereum is lower than that of Bitcoin.

Besides selfish mining and double-spending, blockchains are vulnerable to other types of attacks.
In 2015, Eyal proposed the "miner's dilemma" theory in~\cite{eyal2015miner}.
% By analyzing the competitive relationship between multiple mining pools in Bitcoin in a mining competition from a game theory perspective, 
From a theoretical perspective, this research argued that rational mining pools have an incentive to send members to join competing pools and launch a block withholding attack.
Kwon et al. proposed a novel fork-after-withholding attack in~\cite{kwon2017selfish} and showed that the attack is very profitable and that a large pool can definitely win by launching the attack against a small pool without getting into a miner's dilemma.
Gao et al. investigated two novel attack methods, power-adjusting-withholding (PAW) and bribery-selfish-mining (BSM) in~\cite{gao2019power}, and showed that PAW could evade miners' dilemmas, while BSM increases attackers' gains by 10\% over selfish mining.
Gaži et al. further analyzed the impact of resource centrality on security thresholds in Bitcoin at a theoretical level in~\cite{gavzi2020tight}.

Recently, the security of decentralized governance has attracted a lot of attention.
In~\cite{jeong2020centralized}, Jeong et al. theoretically studied the optimal number of votes per account in DPoS blockchains that employ the approval voting rule.
In~\cite{Defi}, Monday Capital and DappRadar investigated the decentralized governance of six DAOs (Decentralized Autonomous Organizations) where decisions are made through stake-weighted votes and demonstrated that these projects tended to be extremely centralized.
In~\cite{fritsch2022analyzing}, Fritsch et al. empirically studied the distribution of voting power in three prominent DAOs and showed that the governance is dominated by a few voters.
In this paper, inspired by these recent works, we have formally modeled coin-based voting governance, takeover attack/resistance and the \textit{takeover game}. 
Our work demonstrates the theoretical upper bound of active resistance for blockchains that employ different voting rules, and we presented the first large-scale empirical study of the passive takeover resistance of EOSIO, Steem and TRON.

% To sum up, we believe the study of decentralization is developing in three directions, from pool-level to individual-level, from PoW to other consensus protocols, and from snapshot study to longitudinal study. 
% For instance, \cite{kwon2019impossibility} performed pool-level, snapshot study over various blockchains; \cite{zeng2021characterizing} performed individual-level, longitudinal study over Ethereum; our study in this paper performed double-level, longitudinal study over Steem, Hive and Ethereum. 
% We believe the next step is to expand double-level, longitudinal studies to more blockchains. However, this is non-trivial. For instance, \cite{zeng2021characterizing} experienced difficulties in collecting individual-level data in Ethereum, and our study experienced difficulties in reconstructing historical stake snapshots in Steem. 
% We believe our work would potentially facilitate more future works.

%-------------------------------------------------------------------------------
\section{Conclusion}
\label{sec:conclusion}
%-------------------------------------------------------------------------------

In this paper, we demonstrate that the resistance of a DPoS blockchain to takeovers is governed by both the theoretical design and the actual use of its underlying coin-based voting system.
After modeling the coin-based voting system and formalizing the \textcolor{black}{takeover attack and resistance} model, 
we theoretically model a game between an attacker and \textcolor{black}{the cooperative resisters} and demonstrate that the current active takeover resistance is far below the theoretical upper bound.
We then present the first large-scale empirical study of the passive takeover resistance of EOSIO, Steem and TRON.
\textcolor{black}{The results demonstrate the diversity of voter preferences, which significantly \textcolor{black}{affects} the passive takeover resistance when the parameters of the coin-based voting system change.}
Our study suggests potential ways to improve the takeover resistance of DPoS blockchains, including the recommended configuration settings of the system based on our theoretical and empirical analyses and a hybrid approach in which both passive and active resistance are combined to improve takeover resistance.
We believe the study presented in this work provides novel insights into the security of coin-based voting governance and can potentially facilitate more future work on designing new voting rules for decentralized governance that provide more compliance with resistance to takeovers.
\textcolor{black}{
Additionally, we suggest further investigation into a broader range of voting systems (e.g., Single Transferable Vote) could potentially uncover voting methods that improve the security of coin-based voting governance. We also recommend researching other governance models that combine coin with reputation and contribution as the weight for voting, which could potentially improve the overall security and fairness of the governance model.}

% \noindent \textbf{Ethical consideration }
% In this paper, we have only collected public blockchain data and we did not provide any information that may potentially cause harm.
% TRON's takeover of Steem is well-known on the Web. 
% Our work mainly focuses on its modeling and the potential ways of improving the resistance to takeovers.

\section*{Acknowledgments}

We would like to thank the anonymous CCS shepherd and reviewers for their constructive feedback that helped us improve the paper. 
% Chao Li's research is partially supported by the National Key R\&D Program of China under Grant No. 2020YFB1005604 and the Beijing Natural Science Foundation under Grant No. M22039.
% This work was supported in part by the National Natural Science Foundation of China (Grant Nos. 62202038, U21A20463, U22B2027, 62272031), the Beijing Natural Science Foundation (Grant Nos. M23019, 4212008), and the Fundamental Research Funds for the Central Universities of China (Grant Nos. ZG216S2373, 2022JBMC007).
% Balaji Palanisamy's research is partially supported by the US National Science Foundation under Grant \#2020071. 
% Any opinions, findings, and conclusions or recommendations expressed in this material are those of the author(s) and do not necessarily reflect the views of the funding agencies.
% Runhua Xu is supported in part by the ...
% Li Duan is supported in part by the ...
% Jiqiang Liu is supported in part by the ...
% Wei Wang is supported in part by the ...

% Chao Li is supported in part by the Beijing Natural Science Foundation under Grant No. M22039 and the National Natural Science Foundation of China under Grant No. 62202038. 
% Balaji Palanisamy's research is partially supported by the US National Science Foundation under Grant \#2020071. 
% , the National Natural Science Foundation of China under Grant No. 62202038
% , and the Fundamental Research Funds for the Central Universities of China under Grant No. 2022JBMC007. 

\renewcommand\refname{Reference}

\bibliographystyle{plain}
\urlstyle{same}

\bibliography{main.bib}

\appendix

\vspace{30mm}
\section{The call-to-action}
\label{appendix_A}

\textcolor{black}{
On the day of the takeover, a prominent Steem community member posted a call-to-action~\cite{call-to-action}.
As illustrated in Table~\ref{table:cta}, the call-to-action (\#1) attracted a remarkable number of comments (449 comments) within ten days, surpassing the two open letters about the takeover posted by TRON (\#2, \#3) and becoming the most-discussed post during this period.}

\textcolor{black}{
The call-to-action presented two recommendations: }

% \vspace{1mm}
% \begin{mdframed}[innerleftmargin=1.6pt]
%   % \centerline{\textbf{Service setup protocol}}
%   \begin{packed_enum}[leftmargin=*]
%     \item \textcolor{black}{Either PROXY ME (i.e., the author).} 
%     \item \textcolor{black}{Or VOTE HERE: \url{https://steemitwallet.com/~witnesses} \\ 
%     Vote for 22-42 at a minimum, we need to vote for the same witnesses to maximize our votes! Use all 30 of your votes!}
%   \end{packed_enum}
% \end{mdframed}
% \vspace{1mm}

\noindent\fbox{%
\parbox{\dimexpr\linewidth-2\fboxsep-2\fboxrule}{
  \begin{enumerate}[leftmargin=*,label={}]
    \item \textcolor{black}{Either PROXY ME (i.e., the author).} 
    \item \textcolor{black}{Or VOTE HERE: \url{https://steemitwallet.com/~witnesses} \\
    Vote for 22-42 at a minimum, we need to vote for the same witnesses to maximize our votes! Use all 30 of your votes!}
  \end{enumerate}
}}

\noindent \textcolor{black}{In other words, it advised voters to either utilize liquid democracy to delegate voting power to the author of the call-to-action or employ approval voting to cast votes for the 21 BP candidates belonging to the Steem community who were ranked 22-42 at the time.}

\begin{figure}
  \centering
  {
      \includegraphics[width=0.9\columnwidth]{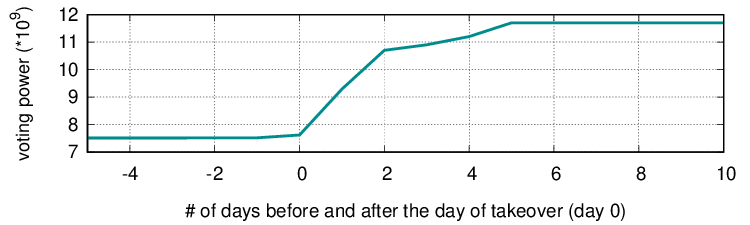}
  }
  \vspace{-3mm}
  \caption {\small The increment of the author's voting weight.}
  \vspace{-3mm}
  \label{fig:dan} 
\end{figure}

\textcolor{black}{
As shown in Figure~\ref{takeover_10days} of Section~\ref{s4.2}, the call-to-action yielded exceptional results, and the recommended BP candidates quickly received substantial voting power support. 
Further investigation revealed that many voters also followed the first suggestion of the call-to-action.
As depicted in Figure~\ref{fig:dan}, they delegated their voting power to the author of the call-to-action, thereby increasing the author's voting weight to 156\% of that prior to the takeover.}

\section{Community-to-community takeover}
\label{appendix_B}

% \noindent \textbf{The impact of complexity.}
In the analysis of Section~\ref{sec:game}, we have made an implicit assumption that $\mathcal{A}$ (or $\mathcal{R}$) is always capable of evenly distributing $\zeta p_a$ (or $\zeta p_r$) across $t$ (or $n-t+1$) candidates.
% , which may be non-trivial in practice.
For instance, in a system that $t=7$ and $v=5$, $\mathcal{A}$ (or $\mathcal{R}$), who owns $\zeta_a p_a=5 \times 70\delta=350\delta$, needs to assign an amount of  $70\delta$ to 5 candidates.
To do that, they need to create 7 accounts as voters, transfer an amount of $10\delta$ voting power to each voter and use each voter to vote for 5 different candidates.
% The procedure thus becomes quite cumbersome, which may be difficult to get properly implemented when $\mathcal{A}$ or $\mathcal{R}$ is formed by a group of entities.
\textcolor{black}{
This process thus becomes highly complex, which may be difficult to implement in practice.
% which may not pose a problem when $\mathcal{A}$ (or $\mathcal{R}$) has a strong control over its voting power (e.g., the TRON founder), but could be difficult to implement when $\mathcal{A}$ (or $\mathcal{R}$) lacks control over its voting power.
}
\textcolor{black}{For example, if Steem's call-to-actions included the aforementioned complex steps, it may be difficult to attract co-resisters who are willing or able to follow them.}

\begin{table}
  \small
  \begin{center}
  \begin{tabular}{|p{0.3cm}|p{5.8cm}|p{1.3cm}|}
  \hline
   \# & {\textbf{post}} & {\textbf{comments}}   \\
      \hline
      1 & call-to-action-earn-upvotes-to-vote-for-witnesses & 449   \\
      \hline % \hdashline
      2 & steemit-witness-voting-policy & 385   \\
      \hline % \hdashline 
      3 & an-open-letter-to-the-community-hf22-5  & 370    \\
      \hline % \hdashline 
      4 & my-resignation-from-steemit  & 182    \\
      \hline % \hdashline 
      5 & an-update  & 139    \\
      % \hline % \hdashline 
      % 6 & quick-update  & 120    \\
      % \hline % \hdashline 
      % spud-x-free-steem  & 116    \\
      \hline
  \end{tabular}
  \end{center}
  % \vspace{-2mm}
  \caption{\small The top 5 posts with the most comments after the takeover.}  
  \vspace{-8mm}    
  \label{table:cta}
\end{table}

\textcolor{black}{
In community-to-community takeovers, the two players, $\mathcal{A}$ and $\mathcal{R}$, might represent two different communities within the same blockchain (e.g., two distinct national communities) or from different blockchains (e.g., members of blockchain A exchanging tokens from blockchain B to attack blockchain B).
In these scenarios, the simplification of the attack and resistance processes, namely call-to-actions, becomes crucial.
Therefore, we assume that both $\mathcal{A}$ and $\mathcal{R}$ are communities and employ a minimum number of \textit{simple} call-to-actions, denoted as $z$, such that $vz \ge t$ for $\mathcal{A}$ and $vz \ge (n-t+1)$ for $\mathcal{R}$.
We can consider each \textit{simple} call-to-action as a pool with an upper limit, which simply accumulates voting power from voters. Once the limit is reached, the pool casts its voting power toward $v$ candidates, and is then replaced by a new pool created by another \textit{simple} call-to-action, which votes for another $v$ candidates.
In this way, $\mathcal{A}$ (or $\mathcal{R}$) could leverage a minimum number of \textit{simple} call-to-actions to vote for each candidate at least once but at most twice.}
Specifically, in an approval voting system, $\mathcal{A}$ would employ only $z=\lceil\frac{t}{v}\rceil$ \textit{simple} call-to-actions and assign an amount of $\frac{p_a}{\lceil\frac{t}{v}\rceil}$ voting power to each candidate.
Similarly, $\mathcal{R}$ would employ $z=\lceil\frac{n-t+1}{v}\rceil$ \textit{simple} call-to-actions and assign an \textcolor{black}{amount} of $\frac{p_r}{\lceil\frac{n-t+1}{v}\rceil}$ voting power to each candidate.
This gives the following theorem:

\begin{theorem}
  \label{theorem:2}
  In an approval-voting supermajority-governing system, if both $\mathcal{A}$ and $\mathcal{R}$ are communities and employ a minimum number of \textit{simple} call-to-actions, the resistance $R_A$ would be upper-bounded by $\lceil\frac{t}{n-t+1}\rceil p_r$.
\end{theorem}

% \begin{lemma}
%   Protocol \textsf{T-Watch} satisfies secrecy.
% \end{lemma}

\begin{proof}
  Based on two properties of ceiling functions~\cite{graham1989concrete}, $P1$
  ($x_1 \le x_2 \Rightarrow \lceil x_1 \rceil \le \lceil x_2 \rceil$) and $P2$
  ($\lceil mx \rceil = \lceil x \rceil+\lceil x-\frac{1}{m} \rceil+\cdots+\lceil x-\frac{m-1}{m} \rceil$ for positive integer $m$),
  we have:
  \begin{align*}
    R_A
    =\frac{\lceil \frac{t}{v}  \rceil}{\lceil \frac{n-t+1}{v} \rceil} p_r
    &=   \frac{\lceil \frac{t}{n-t+1} \cdot \frac{n-t+1}{v} \rceil}{\lceil \frac{n-t+1}{v} \rceil} p_r  \\
    &\le \frac{\lceil \lceil\frac{t}{n-t+1}\rceil \cdot \frac{n-t+1}{v} \rceil}{\lceil \frac{n-t+1}{v} \rceil} p_r 
    \ (based\ on\ P1) \\
    &\le \frac{\lceil\frac{t}{n-t+1}\rceil \cdot \lceil\frac{n-t+1}{v}\rceil}{\lceil \frac{n-t+1}{v} \rceil} p_r
    \ (based\ on\ P2) \\
    &=   \lceil\frac{t}{n-t+1}\rceil p_r
  \end{align*}
\end{proof}

Then, based on Lemma~\ref{lemma:4} and Theorem~\ref{theorem:2}, we can easily prove Lemma~\ref{lemma:5} by injecting $v=n-t+1$ into $R_A=\frac{\lceil \frac{t}{v}  \rceil}{\lceil \frac{n-t+1}{v} \rceil} p_r$.

\begin{table}
  \small
  \begin{center}
  \begin{tabular}{|p{1cm}|p{2cm}|p{2cm}|}
  \hline
  {\textbf{Chain}} & {\textbf{$R_A$ (current)}} & {\textbf{$R_A$ (upper)}}  \\
      \hline
      EOSIO & $p_r$ & $3 p_r$  \\
      \hline % \hdashline
      Steem & $p_r$ & $5 p_r$  \\
      \hline % \hdashline 
      TRON  & $p_r$ & $3 p_r$  \\
      \hline
  \end{tabular}
  \end{center}
  % \vspace{-2mm}
  \caption{\small The current active resistance $R_A$ and the theoretical upper bound of $R_A$ in community-to-community takeovers.}
  %  without (middle column) or with (right column) the `lazy' assumption.}  
  \vspace{-6.0mm}    
  \label{table:revisit-2}
\end{table} 

\begin{lemma}
  \label{lemma:5}
    Given a pair of parameters $(t,n)$, by setting the MaxVote parameter $v=n-t+1$, the active takeover resistance $R_A$ can reach the upper bound whether or not the players are communities that employ a minimum number of \textit{simple} call-to-actions.
\end{lemma}

\textcolor{black}{
  Finally, Table~\ref{table:revisit-2} illustrates the theoretical upper bound of $R_A$ when players are assumed to be communities employing a minimal number of \textit{simple} call-to-actions, by setting $v=n-t+1$. The results indicate that $R_A$ achieves even higher values compared to those presented in Table~\ref{table:revisit}.}

% We now revisit\footnote[9]{For Steem, we ignore the rotational seat and set $n=20$.} the resistance $R_A$ deduced from the parameters of EOS, Steem and TRON shown in Table~\ref{table:summary}.
% % {\bf For Steem, we ignore the rotational seat and set $n=20$.}
% % {\bf Chao, I suggest moving the previous line as a footnote.}
% The results shown in Table~\ref{table:revisit} demonstrate that the current resistance of DPoS blockchains is far below the theoretical upper bound.

% \begin{figure}
% \centering
% {
%     \includegraphics[width=1.0\columnwidth]{./figures/heatmap_afterfork.png}
% }
% \vspace{-6mm}
% \caption {Heatmap of top-50 producers' normalized block creation rates during 48 months in Steem and Ethereum}
% \vspace{-2mm}
% \label{sec3_01} 
% \end{figure}

% \begin{figure}
% \centering
% {
%     \includegraphics[width=1.0\columnwidth]{./figures/heatmap_afterfork_stakeholder.png}
% }
% \vspace{-6mm}
% \caption {Heatmap of top-50 producers' normalized block creation rates during 48 months in Steem and Ethereum}
% \vspace{-2mm}
% \label{sec3_01} 
% \end{figure}

\end{document}